\documentclass{JHEP3}
\usepackage[english]{babel}
\usepackage{cite}

\usepackage{gensymb}

\usepackage{amssymb}
\usepackage{amsmath}
\usepackage{amsfonts}
\usepackage{graphicx}
\usepackage{psfrag}

\def\tr{{\rm tr}}

\newcommand{\Tr}{\hbox{Tr}}

\def\slashchar#1{\setbox0=\hbox{$#1$}
   \dimen0=\wd0
   \setbox1=\hbox{/} \dimen1=\wd1
   \ifdim\dimen0>\dimen1
      \rlap{\hbox to \dimen0{\hfil/\hfil}}
      #1
   \else
      \rlap{\hbox to \dimen1{\hfil$#1$\hfil}}
      /
   \fi}

\def\ps{\slashchar{p}}

\newcommand{\shat}{{\hat s}}

\title{Forward $J/\psi$ and very backward jet inclusive production at the LHC}

\author{R.~Boussarie\\
Institute of Nuclear Physics, Polish Academy of Sciences, Radzikowskiego 152, PL-31-342 Krak\'ow, Poland\\
Email: \email{Renaud.Boussarie@ifj.edu.pl}}
\author{B.~Duclou\'e\\
Department of Physics, University of Jyv\"askyl\"a, P.O. Box 35, 40014 University of Jyv\"askyl\"a, Finland\\
Helsinki Institute of Physics, P.O. Box 64, 00014 University of Helsinki, Finland\\
Email: \email{bertrand.b.ducloue@jyu.fi}}
\author{ L. Szymanowski\\
National Center for Nuclear Research (NCBJ), 00-681 Warsaw, Poland\\
Email: \email{Lech.Szymanowski@ncbj.gov.pl}}
\author{S. Wallon\\
Laboratoire de Physique Th\'{e}orique (UMR 8627), CNRS, Univ. Paris-Sud,
Universit\'{e} Paris-Saclay, 91405 Orsay Cedex, France\\
UPMC, Universit\'{e} Paris 06, Facult\'{e} de Physique, 4 place Jussieu, 75252 Paris, France \\
E-mail: \email{wallon@th.u-psud.fr}}

\abstract{In the spirit of Mueller-Navelet dijet production, we propose and study the inclusive production of a forward $J/\psi$ and a very backward jet at the LHC as an observable to reveal high-energy resummation effects \`a la BFKL. We obtain several predictions, which are based on the various mechanisms discussed in the literature to describe the production of the $J/\psi$, namely, NRQCD singlet and octet contributions, and the color evaporation model.}

\date{\today}
\begin{document}

\pagestyle{empty}
\newpage

\mbox{}

\pagestyle{plain}

\setcounter{page}{1}

\section{Introduction}

The understanding of the 
high energy behaviour of QCD in the perturbative Regge limit remains one of the most important and longstanding theoretical questions in particle physics. In the linear regime where gluonic saturation effects are not expected to be essential, QCD dynamics are described using the BFKL formalism~\cite{Fadin:1975cb,Kuraev:1976ge,Kuraev:1977fs,Balitsky:1978ic}, in the $k_t$-factorization~\cite{Cheng:1970ef, FL, GFL, Catani:1990xk, Catani:1990eg, Collins:1991ty, Levin:1991ry} framework. In order to reveal these resummation effects, first with 
leading logarithmic (LL) precision, and more recently at 
next-to-leading logarithmic (NLL) accuracy,
many processes have been proposed. One of the most promising ones is the inclusive dijet production with a large rapidity
separation, as proposed by Mueller and Navelet~\cite{Mueller:1986ey}. This idea led to many studies, now at the level of NLL precision.

Recent $k_t$-factorization studies of Mueller-Navelet  jets~\cite{Colferai:2010wu,Ducloue:2013hia,Ducloue:2013bva,Ducloue:2014koa,Caporale:2012ih,Caporale:2013uva,Caporale:2014gpa,Celiberto:2015yba} were successful in describing such events at the LHC~\cite{Khachatryan:2016udy}, exhibiting the very first sign of BFKL resummation effects at the LHC. 
To test the universality of such effects,
we propose to apply a similar formalism to study the production of a forward $J/\psi$ meson and a very backward jet with a rapidity interval that is large enough to probe the BFKL dynamics but small enough for
both the $J/\psi$ and the jet to be in the detector acceptance at LHC experiments such as ATLAS or CMS.\footnote{For example, at CMS the CASTOR calorimeter allows one to tag a jet down to $Y_2=-6.55$ in rapidity while the $J/\psi$ could be reconstructed up to $Y_1=2.4$, thus with a maximum interval in rapidity of almost 9, more than sufficient to see BFKL resummation effects.}
Although $J/\psi$ mesons were first observed more than 40 years ago, the theoretical mechanism for their production is still to be fully understood and the validity of some models remains a subject of discussions (for recent reviews see for example refs.~\cite{Brambilla:2010cs,Bodwin:2013nua}).
In addition, most predictions for charmonium production rely on collinear factorization, in which one considers the interaction of two on-shell partons emitted by the incoming hadrons, to produce a charmonium accompanied by a fixed number of partons. On the contrary, in this work the $J/\psi$ meson and the tagged jet are produced by the interaction of two collinear partons, but with the resummation of any number of accompanying unobserved partons, as usual in the $k_t$-factorization approach. 

Here we will compare two different approaches for the description of charmonium production. First we will use the NRQCD formalism~\cite{Bodwin:1994jh}, in which the charmonium wavefunction is expanded as a series in powers of the relative velocity of its constituents. Next we will apply the Color Evaporation Model (CEM), which relies on the local-duality hypothesis~\cite{Fritzsch:1977ay,Halzen:1977rs}. 
Finally we will show numerical estimates of the cross sections and of the azimuthal corrrelations between the $J/\psi$ and the jet obtained in both approaches.
We will rely on the 
Brodsky-Lepage-Mackenzie (BLM)
procedure~\cite{Brodsky:1982gc} to fix the renormalization scale, 
as it was adapted 
to the resummed perturbation theory \`a la BFKL in refs.~\cite{Brodsky:1998kn,Brodsky:2002ka}, which some of us applied to Mueller-Navelet jets in ref.~\cite{Ducloue:2013bva}. Below, we will only discuss in detail the new elements related to the various $J/\psi$ production mechanisms. All details related to the BFKL evolution at NLL can 
be found in refs.~\cite{Colferai:2010wu,Ducloue:2013hia}, while the details related to the application of the BLM scale fixing in our study are presented in ref.~\cite{Ducloue:2013bva}.

\section{Determination of the $J/\psi$ meson vertex}

We start with the determination of a general meson $M$ production vertex (the fact that we will restrict ourselves to $J/\psi$ in the rest of this paper plays no role at this stage). For the moment, we do not consider any specific model for its production. We generically denote with an index $M$ the kinematical variables attached to the system made of the meson and the possible accompanying unobserved particles, and use an index $V$ for the kinematical variables attached to the $J/\psi$ meson itself.

\begin{figure}[t]
\center
\psfrag{p1}{$\hspace{-.4cm} H(p_1)$}
\psfrag{p2}{$\hspace{-.4cm} H(p_2)$}
\psfrag{q1}{$x \, p_1$}
\psfrag{q2}{$\hspace{-.2cm} x' \, p_2$}
\psfrag{k}{$k$}
\psfrag{d}{$k'$}
\psfrag{q}{$p_J$}
\psfrag{a}{$a$}
\psfrag{b}{$b$}
\psfrag{M}{$p_M$}
\psfrag{X}{$X$}
\psfrag{Y}{$Y$}
\raisebox{1cm}{\includegraphics[scale=1]{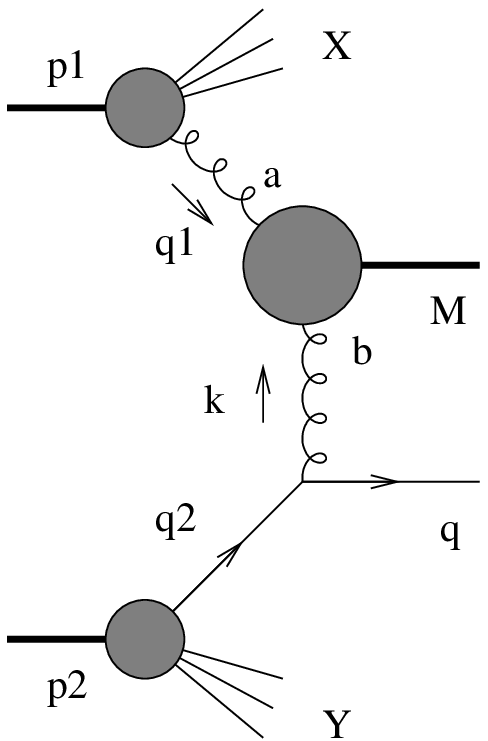}} \hspace{2cm} \raisebox{0cm}{\includegraphics[scale=1]{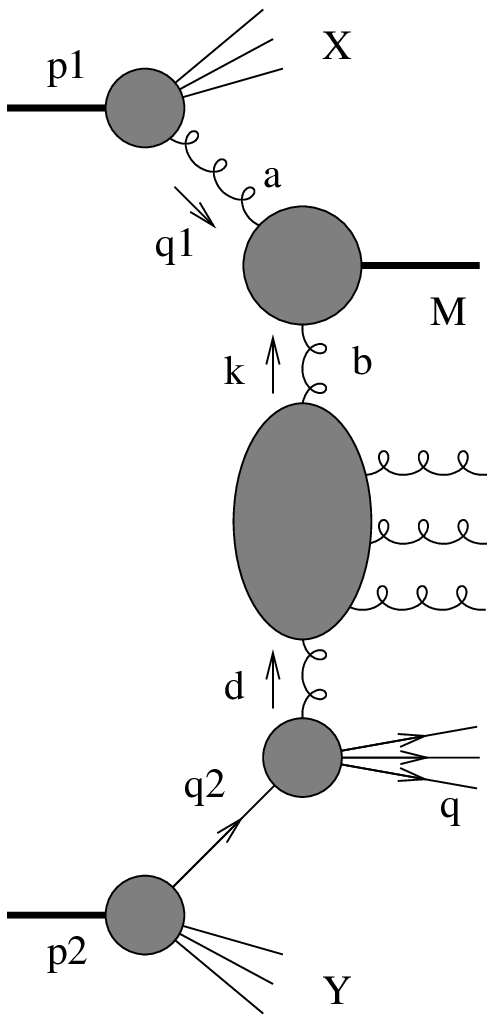}} 
\caption{The high-energy hadroproduction of a meson $M$ and a jet (here originating from a quark) with a large rapidity between them. Left: Born approximation. Right: inclusion of BFKL-like resummation effects due to multiple emissions of gluons and of higher order jet vertex corrections.}
\label{Fig:Process}
\end{figure}

The  inclusive high-energy hadroproduction process of such a meson $M$, via two gluon fusion, with a remnant $X$ and a jet
with a remnant $Y$ separated by a large rapidity difference between the jet and the meson, in scattering of a hadron $H(p_1)$ with a hadron $H(p_2)$, is illustrated in figure~\ref{Fig:Process}, where as a matter of illustration, we consider the parton coming out of the hadron $H(p_1)$ to be a gluon and the parton coming out of the hadron $H(p_2)$ to be a quark. For the sake of illustration, we suppose that the meson is produced in the fragmentation region of the hadron $H(p_1)$, named as forward, while the jet is produced in the fragmentation region of the hadron $H(p_2)$, named as backward. 
On one hand, the longitudinal momentum fractions of the 
jet and of the meson are assumed to be large enough so that the usual collinear factorization applies (the hard scales are provided by the heavy meson mass  and by the transverse momentum of the jet), and we can neglect any transverse momentum, denoting the momentum of the upper (resp. lower) parton as $x \, p_1$ (resp. $x' \, p_2$), their distribution being given by usual parton distribution functions (PDFs). On the other hand, the $t-$channel exchanged momenta (e.g. $k$ in the lhs of figure~\ref{Fig:Process}, or the various ones involved in the rhs of figure~\ref{Fig:Process}) between the meson and the jet cannot be neglected due to their large relative rapidity, and we rely on $k_t-$factorization. 

According to this picture,\footnote{We use the same notations as in refs.\cite{Colferai:2010wu,Ducloue:2013hia}.}
the differential cross section can be written as
\begin{equation}
\frac{d\sigma}{dy_V d|p_{V\bot}|d\phi_V  dy_{J} d|p_{J\bot}|d\phi_{J}} 
= \sum_{{a, b}} \int_0^1 \!dx \int_0^1 \!dx' f_{a}(x) f_{b}(x') \frac{ d \hat{\sigma}}{dy_V d|p_{V\bot}|d\phi_V  dy_{J} d|p_{J\bot}|d\phi_{J}} \, ,
\end{equation}
where $f_{a, b}$ are the standard parton distribution functions of a parton $a (b)$ in the according hadron. 

In $k_t$-factorization, the partonic cross section reads
\begin{equation}
\frac{ d \hat{\sigma}}{dy_V d|p_{V\bot}|d\phi_V  dy_{J} d|p_{J\bot}|d\phi_{J}}
= \int d^2 k_\perp  \, d^2 k_\perp' V_{V,a}(k_\perp,x) \, G(-k_\perp,-k_\perp',\shat)\,V_{J, b}(-k_\perp',x') \, ,
\label{eq:bfklpartonic}
\end{equation}
where $G$ is the BFKL Green's function depending on $\shat=x x' s$, 
denoting as $\sqrt{s}$  the center-of-mass energy  of the two colliding hadrons.

At leading order (LO), the jet vertex reads~\cite{Bartels:2001ge,Bartels:2002yj}:
\begin{align}
V_{J, a}^{(0)}(k_\perp,x) = & h_{a}^{(0)}(k_\perp)\mathcal{S}_J^{(2)}(k_\perp;x) \, , \label{def:V0} \\ h_{a}^{(0)}(k_\perp) = & \frac{\alpha_s}{\sqrt{2}}\frac{C_{A/F}}{k_\perp^2} \, , \quad \mathcal{S}_J^{(2)}(k_\perp;x) = \delta\left(1-\frac{x_J}{x}\right)|p_{J \perp}|\delta^{(2)}(k_\perp-p_{J \perp}) \, . \label{def:S}
\end{align}
In the definition of $h_{\rm a}^{(0)}$, $C_A=N_c=3$ is to be used for an initial gluon and $C_F=(N_c^2-1)/(2N_c)=4/3$ for an initial quark. Following the notations of refs.~\cite{Bartels:2001ge,Bartels:2002yj}, the dependence of $V$ on the jet variables is implicit.
At next-to-leading order (NLO), the jet can be made of either a single or two partons. The explicit form of these jet vertices can be found in ref.\cite{Colferai:2010wu} as extracted from refs.~\cite{Bartels:2001ge,Bartels:2002yj} after correcting a few misprints of ref.~\cite{Bartels:2001ge}.

The explicit form of the BFKL Green's function $G$, as obtained  at LL~\cite{Fadin:1975cb,Kuraev:1976ge,Kuraev:1977fs,Balitsky:1978ic} and at NLL~\cite{Fadin:1998py,Ciafaloni:1998gs} accuracy, can be found in ref.\cite{Colferai:2010wu}, and will not be reproduced here.

In the rest of the present paper, we will only focus on the case where the meson vertex is treated at lowest order, while the Green's function and the jet vertex will
be treated at NLL. The computation of the NLO $J/\psi$ vertex, which is a quite involved task, is left for further studies.

\begin{figure}[t]
\center
\psfrag{p1}{$\hspace{-.6cm} H(p_1)$}
\psfrag{p2}{$\hspace{-.6cm} H(p_2)$}
\psfrag{q1}{$\hspace{-.2cm} x \, p_1$}
\psfrag{q2}{$\hspace{-.2cm} x' \, p_2$}
\psfrag{k}{$k$}
\psfrag{q}{$p_J$}
\psfrag{a}{$a$}
\psfrag{b}{$b$}
\psfrag{M}{$\!p_M$}
\psfrag{X}{$X$}
\psfrag{Y}{$Y$}
\psfrag{g}{$g_\perp$}
\psfrag{p1s}{$\ps_1$}
\psfrag{p2s}{$\ps_2$}
\hspace{.2cm}\raisebox{2cm}{\includegraphics[scale=.75]{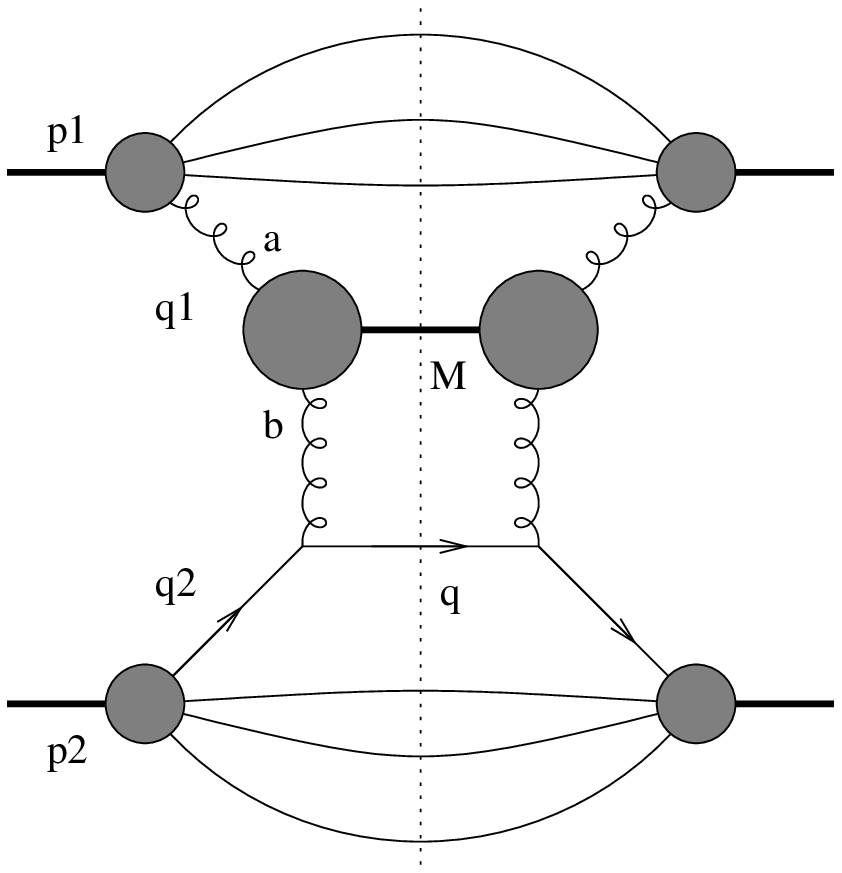}} \quad \psfrag{M}{\raisebox{-.05cm}{$\!M$}}
\psfrag{J}{$\!\!J$}
\includegraphics[scale=.85]{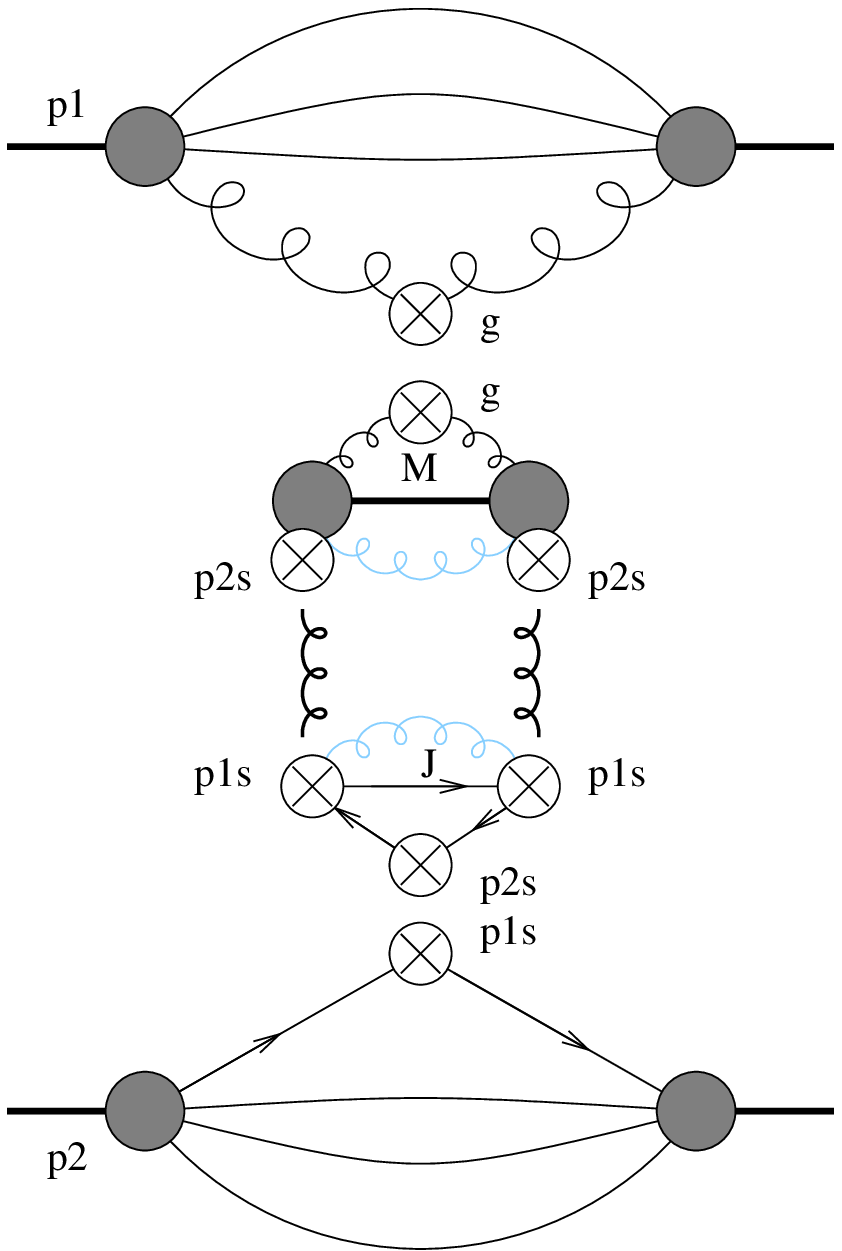}
\caption{Left: square of the amplitude of the Born process. Right: symbolic factorized form of this squared amplitude involving, from top to bottom, the gluonic PDF, the impact factor describing the $g\, g  \to M$ transition, the $t-$channel exchange of two off-shell gluons (in bold), the  vertex describing the $g q \to {\rm jet}$ transition, and the quark PDF. The crosses symbolically denote the appropriate Fierz structure in Lorentz space. Namely, from top to bottom, $g_\perp$ tensors for gluons, due to the collinear factorization of the gluon out of the upper PDF, $\ps_1$ and $\ps_2$ arising from the {\it non-sense} polarizations of the $t-$channel gluons in $k_t$-factorization, and finally $\ps_2$ and $\ps_1$ due to the collinear factorization of the quark out of the down PDF. The black fermions and curved gluon lines symbolize the trace over color and Lorentz indices after the use of the Fierz identity in these two spaces, while the blue (grey in printed black and white) gluons are traces over color 
after the 
use of the Fierz identity in 
color space.}
\label{Fig:Born-Process-factorized}
\end{figure}

To properly fix the normalization, let us focus for a moment on the Born approximation, see the lhs of figure~\ref{Fig:Process}. Then, each building block in the factorized formula (\ref{eq:bfklpartonic}) is treated at lowest order.
In this limit,
our normalizations are  such that the Born Green's function is
\begin{equation}
G^{\rm Born}(k_\perp,k_\perp',\shat)=\delta^2(k_\perp - k_\perp')\,,
\label{G:Born}
\end{equation}
while the jet vertices are given by eqs.~(\ref{def:V0}, \ref{def:S}).
As explained above, the relevant components of the involved momenta read
\begin{equation}
k = \beta p_2 +k_\bot, \quad p_J = x' p_2 +p_{J\bot}, \quad p_M= x \, p_1 + p_{M\bot} ,
\label{Sudakov_k}
\end{equation}
where $k$ is the
$t-$channel 
exchanged momentum.
In the high-energy limit, the ${\cal T}_{Mq}$ -matrix reads
\begin{equation}
{\cal T}_{Mq} = \frac{1}{i} \, \frac{2}{s} \frac{(-i)}{p_{J\bot}^2}  \langle X| A^a_\mu(0)|H(p_1)\rangle g^{\mu \nu}_\bot  {\cal A}_\nu^{ab}\,\bar u(p_J)(-ig \hat p_1 t^b) \langle Y| q(0)|H(p_2)\rangle\;,
\label{Mprocess}
\end{equation}
where $a$ is the color index of a collinear gluon from the hadron $H(p_1)$ and $b$ is the color index of the exchanged $t-$channel gluon. Here ${\cal A}_\nu^{ab}$ denotes the $S$-matrix element describing the $g g \to M$ transition. Its computation will be discussed in detail in the following subsections.
After factorization, illustrated symbolically by figure~\ref{Fig:Born-Process-factorized},
we get
\begin{align}
\sum_{Mq} {\cal T}_{Mq}{\cal T}_{Mq}^* =& \frac{4}{s(p^2_{J\bot})^2} \frac{1}{4N(N^2-1)}\langle H(p_1)|A^{a'}_{\mu' \bot}(0)|X\rangle g_\bot^{\mu' \nu'}
\langle X| A^{a' }_{\nu' \bot}(0)|H(p_1) \rangle \nonumber \\
& \times \sum_M  {\cal A}_{\mu \bot}^{ab}  g^{\mu\nu}_\bot  ( {\cal A}^{ab}_{\nu\bot})^*
g^2 \beta_{J} \langle H(p_2)| \bar q^c(0)|Y\rangle \langle Y| \hat p_1 q^c(0)|H(p_2) \rangle \,. \label{Tsquare}
\end{align}
The phase space measure reads\footnote{This should be understood in an extended way, in particular due to the fact that $M$ might involve several particles, as it is the case for the color singlet NRQCD contribution.}
\begin{align}
d\Phi =& (2\pi )^4 \delta(p_1 + p_2 - [p_X] - [p_Y] - [p_M] - p_{J} ) \nonumber \\
&\times \left[  \frac{d^3p_X}{(2\pi )^32E_X} \right] \left[  \frac{d^3p_Y}{(2\pi )^32E_Y} \right] \left[  \frac{d^3p_M}{(2\pi )^32E_M} \right] \frac{d^3p_{J}}{(2\pi)^32E_{J}} \, . \label{PhSp}
\end{align}
It can be written in a factorized form in terms of the rapidity $y_{J}$ of the quark jet and its transverse momentum $p_{J\bot }$:
\begin{align}
&d\Phi = \nonumber \\
& \frac{2\pi }{s} \int d^2k_\bot \delta^2(-[p_{M\bot}] + k_\bot) \,dx\, \delta(x- [\alpha_M]) \,\delta(1-x -[\alpha_X]) \left[  \frac{d^3p_X}{(2\pi )^32E_X} \right] \left[  \frac{d^3p_M}{(2\pi )^32E_M} \right]
\nonumber \\
& \hspace{0.8cm} \times \delta^2(k_\bot + p_{J\bot}) \, dx' \, \delta(x' - \beta_{J})\,\delta(1-x' - [\beta_Y]) \,\left[  \frac{d^3p_Y}{(2\pi )^32E_Y} \right] \, dy_{J}\, d^2p_{J\bot}\,. \label{PhSpFact}
\end{align}
This $k_t$-factorization formula involves an integration over the transverse momentum $k_t$ of the four-momentum transfer $k$ in the $t-$channel between both vertices.
Using the expressions of the unpolarized quark PDF
\begin{equation}
H^q(x')= \frac{1}{s} \int\, \left[  \frac{d^3p_Y}{(2\pi )^32E_Y} \right] \delta(1-x' - [\beta_Y]) \langle H(p_2)| \bar q(0)|Y\rangle \langle Y| \hat p_1 q(0)|H(p_2)\rangle \, ,
\label{Hq}
\end{equation}
and of the unpolarized gluon PDF,
\begin{equation}
\frac{g(x)}{x}  = - \int  \!  \left[  \frac{d^3p_X}{(2\pi )^32E_X} \right]        \delta(1-x -[\alpha_X]) \langle H(p_1)|A^{a'}_{\mu' \bot}(0)|X\rangle g_\bot^{\mu' \nu'}
\langle X| A^{a' }_{\nu' \bot}(0)|H(p_1) \rangle \, ,
\label{g}
\end{equation}
we obtain an expression for the differential cross section 
\begin{align}
\frac{d \sigma}{dy_{J}d|p_{J\bot}|d\phi_{J}}=\int & \, dx \, g(x)\, dx' \,H^q(x')\,d^2k_\bot\,\delta(x-[\alpha_M])\,\delta^2(k_\bot - [p_{M\bot}]) \left[  \frac{d^3p_M}{(2\pi )^32E_M} \right] \nonumber \\
& \times \frac{8\sqrt{2}\pi^2}{s^2(N^2-1)^2 \,x\,k^2_\bot}  \,  \sum_{[M]}  {\cal A}_{\mu \bot}^{ab} g^{\mu \nu}_\bot   ({\cal A}^{ab}_{\nu\bot})^*  \,\,V_{J,q}^{(0)}(-k_\bot,x')\,, \label{CSecM}
\end{align} 
in which we factorized out the vertex for quark jet production in the Born approximation,
\begin{equation}
V_{J,q}^{(0)}(k_\bot,x') = \frac{g^2}{4\pi \sqrt{2}} \frac{C_F}{|k_\bot|} \,\delta\left(1-\frac{x_J}{x'}\right) \,\delta^2(k_\bot - p_{J\bot})\,,
\label{Vq}
\end{equation}
in accordance with eqs.~(\ref{def:V0}, \ref{def:S}).

\subsection{Color-singlet NRQCD contribution}

In the color-singlet contribution  
the system $[M]$ is made of the produced $J/\psi$ charmonium and of the unobserved gluon produced simultaneously with the charmonium in gluon-gluon fusion due to the negative charge-parity of the $J/\psi$. 
We parametrize the momentum $p_V$ of the $J/\psi$ and the momentum $l$
of the unobserved gluon in terms of Sudakov variables, as
\begin{equation}
p_V=\alpha_V p_1 + \frac{M^2_{J/\psi}-p^2_{V\bot}}{\alpha_V s}p_2 +p_{V\bot}\;,\;\;\;\;\;l=\alpha_l \, p_1 - \frac{l_\bot^2}{\alpha_l s} p_2+l_\bot \,.
\label{CSSudakov}
\end{equation}
Thus the expression of
\begin{align}
&\delta(x-[\alpha_M])\,\delta^2(k_\bot - [p_{M\bot}]) \left[  \frac{d^3p_M}{(2\pi )^32E_M} \right] \nonumber \\
& = \delta(x - \alpha_l -\alpha_V)\, \delta^2(k_\bot - l_\bot - p_{V\bot}) 
\frac{d^3l}{(2\pi )^32E_l}\frac{d^3p_V}{(2\pi )^32E_V}
\nonumber \\
& = \frac{1}{4(2\pi )^6}\delta(x-\alpha_l - \alpha_V)\delta^2(k_\bot - l_\bot - p_{V\bot}) \frac{d\alpha_l\,\theta(\alpha_l)}{\alpha_l}\,d^2l_\bot\,dy_V d^2p_{V\bot} \label{CSM}
\end{align}
permits, with the use of (\ref{CSecM}), to write the differential cross section in the form
\begin{align}
\frac{d\sigma}{dy_V d|p_{V\bot}|d\phi_V  dy_{J} d|p_{J\bot}|d\phi_{J}} = \int & dx \, g(x) \,dy \, H^q(y) \, d^2k_\bot \frac{|p_{V\bot}|\sqrt{2} }{2^5 \pi^4 s^2 (N^2-1)^2 k_\bot^2\, x} \frac{\theta(x-\alpha_V)}{x-\alpha_V} \nonumber \\
& \times \sum_{\lambda_V, \lambda_l }  {\cal A}_{\mu \bot}^{ab} g_\bot^{\mu \nu}({\cal A}_{\nu \bot}^{ab})^*  \,\,V_q^{(0)}(-k_\bot,y) \, , \label{CSCrossS}
\end{align}
from which we read off the $J/\psi$ production vertex of the color singlet NRQCD contribution as
\begin{equation}
V_{J/\psi}^{(1)} =\frac{|p_{V\bot}|\sqrt{2} }{2^5 \pi^4 s^2 (N^2-1)^2 k_\bot^2\, x} \frac{\theta(x-\alpha_V)}{x-\alpha_V} \sum_{\lambda_V, \lambda_l }  {\cal A}_{\mu \bot}^{ab} g_\bot^{\mu \nu}({\cal A}_{\nu \bot}^{ab})^* \,.
\label{CSVertex}
\end{equation}
One should note that the above expressions include an integration over the phase space of the unobserved gluon with momentum $l\,.$ 
The vertex which allows to pass from open $q \bar{q}$ production to $J/\psi$ production in color singlet NRQCD reads~\cite{Guberina:1980dc,Baier:1983va}
\begin{equation}
[v(q)\bar u(q)]^{ij}_{\alpha \beta} \rightarrow \frac{\delta^{ij}}{4N} \left(  \frac{\langle  {\cal O}_1 \rangle_V}{m} \right)^{1/2} \left[ \hat \epsilon^*_V \left(  2\hat q +2m \right)\right]_{\alpha \beta} ,
\label{CSvertex}
\end{equation}
with the momentum $q = \frac{1}{2}p_V$, $m$ being the mass of the charm quark, $M_{J/\psi}=2m$. In the following we will use the non-perturbative coefficient $C_1$ defined as
\begin{equation}
C_1 \equiv \left(  \frac{\langle  {\cal O}_1 \rangle_V}{m} \right)^{1/2}.
\label{C}
\end{equation}
 The  
matrix element $\langle O_1
\rangle_V$ in NRQCD is related to the
leptonic meson decay rate by~\cite{Bodwin:1994jh}
\begin{equation}
\Gamma[V\to l^+l^-]=\frac{2e_c^2\pi\alpha^2}{3}
\frac{\langle {\cal O}_1\rangle_V }{m^2} 
\left( 1-\frac{16\alpha_s}{3\pi}\right) .
\label{decay}
\end{equation}
Here $\alpha$ is the fine-structure constant and $e_c=2/3$ is the electric charge of the charm quark. Equation (\ref{decay}) includes the one-loop
QCD correction \cite{Barbieri:1975ki,Billoire:1977mp,Celmaster:1978yz} and $\alpha_s$ is the strong coupling
constant. One can use the value of this decay rate to fix $\langle {\cal O}_1\rangle_V$ through this relation.
Namely, using the values $\Gamma_{e^+ e^-}= 5.55 \times 10^{-6}$ GeV~\cite{Olive:2016xmw}, $m=1.5$ GeV and a three-loop running coupling with $\Lambda_4=0.305$ GeV, we obtain $\langle{\cal O}_1 \rangle_{J/\psi}=0.444$ GeV$^3$. As quoted in ref.~\cite{Bain:2017wvk}, recent phenomenological analyses~\cite{Butenschoen:2011yh,Chao:2012iv,Bodwin:2014gia} have used slightly smaller values of either 0.387 or 0.440 GeV$^3$, as obtained in refs.~\cite{Eichten:1995ch} and~\cite{Bodwin:2007fz} respectively. In order not to underestimate the uncertainty, in the following we will vary $\langle{\cal O}_1 \rangle_{J/\psi}$ between 0.387 and 0.444 GeV$^3$.

The momentum transfer $k$ in the $t-$channel entering the charmonium vertex has the approximate  form given by eq.~(\ref{Sudakov_k}).
The momentum conservation in the charmonium vertex $xp_1 +k =p_V+l$ leads to the following relations between the Sudakov variables of momenta:
\begin{equation}
x=\alpha_V + \alpha_l\;,\;\;\;\;\;\;\;\; k_\bot=p_{V\bot} + l_\bot \;,\;\;\;\;\; \beta = \frac{4m^2 -p_{V\bot}^2}{\alpha_Vs} - \frac{l_\bot^2}{\alpha_l s} \;.
\label{CSmomcons}
\end{equation}

\def\sca{1.3}
\def\spa{.67cm}
\begin{figure}[t]
\center
\psfrag{a}{}
\psfrag{b}{}
\psfrag{u}{}
\psfrag{d}{}
\psfrag{d}{}
\psfrag{l}{$\ell$}
\psfrag{qu}{$q$}
\psfrag{qd}{$-q$}
\psfrag{p1}[R]{$xp_1$}
\psfrag{p2}[R]{$\beta p_2 + k_\perp \hspace{.6cm}$}
\hspace*{0.05cm}
\begin{tabular}{ccc}
\psfrag{i}{}
\hspace{\spa}\includegraphics[scale=\sca]{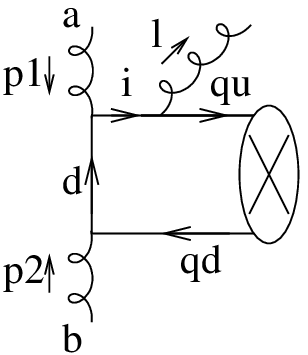}
&
\psfrag{qu}{$\hspace{-.4cm}q$}
\psfrag{qd}{$\hspace{-.7cm}-q$}
\psfrag{i}{}
\hspace{\spa}\raisebox{0cm}{\includegraphics[scale=\sca]{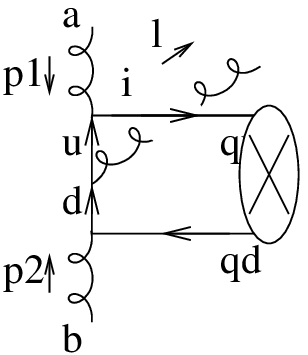}}
&
\psfrag{i}{}
\psfrag{qu}{$\hspace{-.4cm}q$}
\psfrag{qd}{$\hspace{-.2cm}-q$}
\hspace{\spa}\raisebox{0cm}{\includegraphics[scale=\sca]{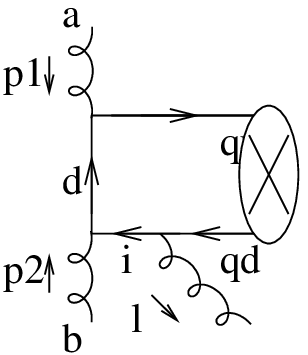}}
\\
$D_1$ & $D_2$ & $D_3$
\\
\\
\psfrag{qu}{$\hspace{-.5cm}-q$}
\psfrag{qd}{$q$}
\psfrag{i}{}
\hspace{\spa}\includegraphics[scale=\sca]{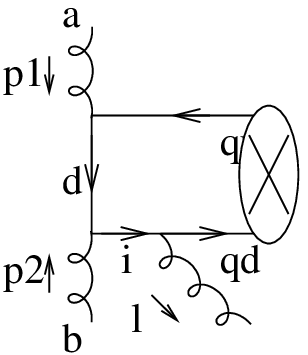}
&
\psfrag{qu}{$\hspace{-.5cm}-q$}
\psfrag{qd}{$\hspace{-.2cm}q$}
\psfrag{i}{}
\hspace{\spa}\raisebox{0cm}{\includegraphics[scale=\sca]{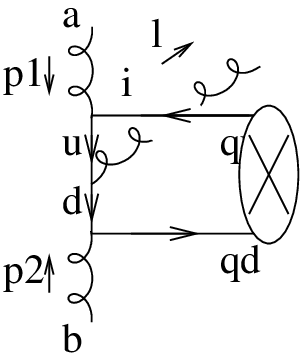}}
&
\psfrag{qu}{$\hspace{-.5cm}-q$}
\psfrag{qd}{$\hspace{-.2cm}q$}
\psfrag{i}{}
\hspace{\spa}\raisebox{0cm}{\includegraphics[scale=\sca]{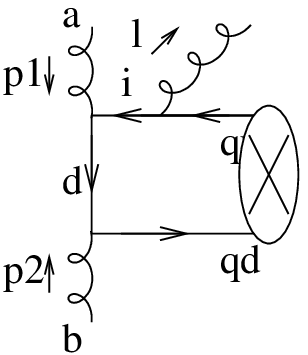}}\\
$D_4$ & $D_5$ & $D_6$
\end{tabular}
\caption{The 6 diagrams contributing to the amplitude in color singlet NRQCD. The blobs with a cross symbolize the Fierz structure of eq.~(\ref{CSvertex}).}
\label{Fig:Diagrams-CE}
\end{figure}
The contribution to the hard part is given by the 6 diagrams shown in figure~\ref{Fig:Diagrams-CE}, which leads to the expressions
\begin{equation}
D_1 = \frac{(-ig)^3i^2C_1}{4N} \tr_c(t^l t^a t^b) \Tr \! \left[ \hat \epsilon^*_V (2\hat q \!+\! 2m) \hat \epsilon^*(l) \frac{\hat q \!+\! \hat l \!+\! m}{(q \!+\! l)^2 \!-\! m^2} \gamma_{\bot}^\mu \frac{\beta \hat p_2 \!+\! \hat k_\bot \!-\! \hat q \!+\! m}{(\beta p_2 \!+\! k_\bot \!-\! q)^2 \!-\! m^2} \hat p_2\right], \hspace{0.45cm}
\label{D1}
\end{equation}
\begin{equation}
D_2 = \frac{(-ig)^3i^2C_1}{4N} \tr_c(t^a t^l t^b) \Tr \! \left[ \hat \epsilon^*_V (2\hat q \!+\! 2m) \gamma_\bot^\mu  \frac{\hat q \!-\! x\hat p_1 \!+\! m}{(q \!-\! xp_1)^2 \!-\! m^2} \hat \epsilon^*(l) \frac{\beta \hat p_2 \!+\! \hat k_\bot \!-\! \hat q \!+\! m}{(\beta p_2 \!+\! k_\bot \!-\! q)^2 \!-\! m^2} \hat p_2\right],
\label{D2}
\end{equation}
\begin{equation}
D_3 = \frac{(-ig)^3i^2C_1}{4N} \tr_c(t^a t^b t^l) \Tr \! \left[ \hat \epsilon^*_V (2\hat q \!+\! 2m) \gamma_\bot^\mu  \frac{\hat q \!-\! x\hat p_1 \!+\! m}{(q \!-\! xp_1)^2 \!-\! m^2} \hat p_2 \frac{- \hat q \!-\! \hat l \!+\! m}{( -q \!-\! l )^2 \!-\! m^2} \hat \epsilon^*(l)\right], \hspace{0.95cm}
\label{D3}
\end{equation}
\begin{equation}
\hspace{0.14cm} D_4 = \frac{(-ig)^3i^2C_1}{4N} \tr_c(t^l t^b t^a) \Tr \! \left[ \hat \epsilon^*_V (2\hat q \!+\! 2m) \hat \epsilon^*(l) \frac{\hat q \!+\! \hat l \!+\! m}{(q \!+\! l)^2 \!-\! m^2} \hat p_2 \frac{x \hat p_1 \!-\! \hat q \!+\! m}{(x p_1 \!-\! q)^2 \!-\! m^2} \gamma_\bot^\mu\right],
\label{D4}
\end{equation}
\begin{equation}
D_5 = \frac{(-ig)^3i^2C_1}{4N} \tr_c(t^b t^l t^a) \Tr \! \left[ \hat \epsilon^*_V (2\hat q \!+\! 2m) \hat p_2  \frac{\hat q \!-\! \beta \hat p_2 \!-\!\hat k_\bot \!+\! m}{(q \!-\! \beta p_2 \!-\! k_\bot)^2 \!-\! m^2} \hat \epsilon^*(l) \frac{ x \hat p_1 \!-\! \hat q \!+\! m}{(x p_1 \!-\! q)^2 \!-\! m^2} \gamma_\bot^\mu\right],
\label{D5}
\end{equation}
\begin{equation}
D_6 = \frac{(-ig)^3i^2C_1}{4N} tr_c(t^b t^a t^l) \Tr \! \left[ \hat \epsilon^*_V (2\hat q \!+\! 2m) \hat p_2  \frac{\hat q \!-\! \beta \hat p_2 \!-\! \hat k_\bot \!+\! m}{(q \!-\! \beta p_2 \!-\! k_\bot)^2 \!-\! m^2} \gamma_\bot^\mu \frac{- \hat q \!-\! \hat l \!+\! m}{( - q \!-\! l)^2 \!-\! m^2} \hat \epsilon^*(l)\right], \hspace{0.15cm}
\label{D6} 
\end{equation}
where $\tr_c$ and $\Tr$ denote respectively the color and the Dirac traces.
Let us observe the following relations between the Dirac traces  ${\Tr}_{D(i)}$  of diagrams $D(i)$ due to charge conjugation invariance:
\begin{equation}
\Tr_{D(1)}  = \Tr_{D(6)}\,,\;\;\;\;    \Tr_{D(2)}    = \Tr_{D(5)}\,,\;\;\;\;    \Tr_{D(3)}  =  \Tr_{D(4)} \, .
\label{Drelations}
\end{equation} 
Consider the color factor: the symmetry property (\ref{Drelations}) results in the appearance in the sum of all diagrams of the symmetric structure constants $d^{ckl}$ of the $SU(N)$ color group only. Thus, we obtain
\begin{align}
{\cal A}^{ab}_{\mu\bot}=&\sum\limits_{i=1}^6 D(i) =  \frac{(-ig)^3i^2C_1}{4N} \frac{d^{abl}}{4}  \left\{     
\frac{2\,\Tr_{D(1)\mu\bot}}{[ (q+l)^2 -m^2][ (\beta p_2 +k_\bot -q)^2 - m^2]} \right. \nonumber \\
& \hspace{0.6cm} \left. +  \frac{2\,\Tr_{D(2)\mu\bot}}{[ (q- x p_1)^2 -m^2   ][ (\beta p_2 +k_\bot -q)^2 - m^2    ]}  + 
\frac{2\Tr_{D(3)\mu\bot}}{[ (q- x p_1)^2 -m^2   ][ (l +q)^2 - m^2    ]}  \right\} \nonumber \\
\equiv & \;  \frac{(-ig)^3i^2C_1}{4N} \frac{d^{abl}}{4} {\cal D}_{\mu\bot}^{\nu \rho} \epsilon^*_{V\rho }(2q)\epsilon^*_\nu (l) \, , \label{TrD}
\end{align}
where we introduced the shorthand notation ${\cal D}_{\mu\bot}^{\nu \rho} \epsilon^*_{V\rho }(2q)\epsilon^*_\nu (l)$ for the sum of all six diagrams contributing to $J/\psi$ production within the color singlet mechanism. One can check that this sum vanishes in the limit 
$k_\perp \to 0,$ as it should be the case for an impact factor in $k_t-$factorization due to its gauge invariance.
For the gluon $g(l)$, we choose the gauge\footnote{Note that the sum of diagrams in this color singlet mechanism is gauge invariant, although the $t-$channel gluon is off-shell: indeed due to the simple single color structure $d^{abl}$ which factorizes, they are QED like.} 
\begin{equation}
p_2\cdot \epsilon^*(l)=0\,,
\label{gaugel}
\end{equation}
which is a natural choice for a meson emitted in the fragmentation region of the hadron of momentum $p_1.$ 
The three different traces then read
\begin{align}
\Tr_{D(1)}^{\,\mu\bot} =& 2m\,\Tr\left[  m^2 \hat \epsilon^*_V \hat \epsilon^*(l) \gamma_\bot^\mu \hat p_2   
+ \hat \epsilon^*_V \hat \epsilon^*(l) (\hat q + \hat l)\gamma_\bot^\mu (  \hat k_\bot - \hat q  )\hat p_2  \right.
\nonumber \\
& \hspace{1.2cm} \left.
+ \hat \epsilon^*_V   \hat q   \hat \epsilon^*(l) (\hat q + \hat l)\gamma_\bot^\mu \hat p_2
+ \hat \epsilon^*_V   \hat q   \hat \epsilon^*(l)  \gamma_\bot^\mu     (  \hat k_\bot   -\hat q ) \hat p_2
\right] \nonumber \\
=&8m\left[   k_\bot^\mu \left(   2 \epsilon^*(l)\cdot q \,\epsilon^*_V\cdot p_2  + \epsilon^*_V\cdot \epsilon^*(l)\, p_2\cdot l    \right) 
- \epsilon^{*\mu}_{V\bot}\left(   k_\bot\cdot \epsilon^*(l)\,p_2\cdot l +4p_2\cdot q\,q\cdot \epsilon^*(l)    \right) \right.
\nonumber \\
& \hspace{0.7cm} \left. +l_\bot^\mu \left(   p_2\cdot \epsilon^*_V\, k_\bot\cdot \epsilon^*(l) -2p_2\cdot q\, \epsilon^*_V\cdot \epsilon^*(l)   \right)\right. \nonumber \\
& \hspace{0.7cm} \left.+\epsilon^{*\mu}_\bot(l) \left(  -k_\bot\cdot l_\bot\, p_2\cdot \epsilon^*_V + k_\bot\cdot \epsilon^*_V\, p_2\cdot l +2p_2\cdot q \,  \epsilon^*_V \cdot l    \right)
\right], \label{TrD1}
\end{align}
\begin{align}
&\Tr_{D(2)}^{\,\mu\bot} = 2m\,\Tr\left[  m^2 \gamma_\bot^\mu \hat \epsilon^*(l) \hat p_2 \hat \epsilon^*_V    
+ \hat \epsilon^*_V   \hat q \gamma_\bot^\mu  \hat \epsilon^*(l) ( \hat k_\bot -   \hat q ) \hat p_2  \right.
\nonumber \\
& \hspace{2.7cm} \left.
+ \hat \epsilon^*_V   \hat q  \gamma_\bot^\mu (\hat q - x \hat p_1)  \hat \epsilon^*(l)  \hat p_2
+ \hat \epsilon^*_V   \gamma_\bot^\mu (\hat q - x \hat p_1)  \hat \epsilon^*(l)    (  \hat k_\bot   -\hat q ) \hat p_2
\right] \nonumber \\
& = 2m \!\left\{
8q_\bot^\mu \left(    -2p_2\cdot q\, \epsilon^*_V\cdot \epsilon^*(l) + \epsilon^*(l)\cdot k_\bot\,p_2\cdot \epsilon^*_V   \right)
+8xp_2\cdot q\,\left(  \epsilon^{*\mu}_{V\bot}\,p_1\cdot \epsilon^*(l) - \epsilon^{*\mu}_\bot(l)\,p_1\cdot \epsilon^*_V      \right) \right.
\nonumber \\
& \hspace{1.1cm} \left.
+x\left[   -2s \epsilon^{*\mu}_{V\bot}\, k_\bot\cdot \epsilon^*(l)  -2s \epsilon^{*\mu}_\bot(l)\,k_\bot\cdot \epsilon^*_V   + k_\bot^\mu \left(   2s \epsilon^*_V\cdot \epsilon^*(l) -4 p_2\cdot \epsilon^*_V \, p_1\cdot \epsilon^*(l)     \right)\right]
\right\}, \label{TrD2}
\end{align}
and
\begin{align}
\Tr_{D(3)}^{\,\mu\bot} =& 2m\,\Tr\left[  m^2 \hat \epsilon^*_V \gamma_\bot^\mu  \hat p_2 \hat \epsilon^*(l)    
- \hat \epsilon^*_V   \gamma_\bot^\mu (\hat q - x \hat p_1)  \hat p_2  (\hat q +\hat l) \hat \epsilon^*(l)   \right.
\nonumber \\
& \hspace{1.2cm} \left.
- \hat \epsilon^*_V   \hat q  \gamma_\bot^\mu \hat p_2    (\hat q + \hat l)  \hat \epsilon^*(l)  
+ \hat \epsilon^*_V  \hat q  \gamma_\bot^\mu (\hat q - x \hat p_1) \hat p_2  \hat \epsilon^*(l)  
\right] \nonumber \\
=&2m\left\{
8q^\mu_\bot \left( -2 q\cdot \epsilon^*(l)\, p_2\cdot \epsilon^*_V +\epsilon^*_V\cdot l\, p_2\cdot \epsilon^*(l) -p_2\cdot l\, \epsilon^*_V\cdot \epsilon^*(l)   \right) \right. \nonumber \\
& \hspace{0.8cm} + x \left[
4s \epsilon^{*\mu}_{V\bot} \, q\cdot \epsilon^*(l) \left. + 2s \left( - \epsilon^{*\mu}_\bot(l) \, l\cdot \epsilon^*_V +l^\mu_\bot\, \epsilon^*(l)\cdot \epsilon^*_V  \right) -4l_\bot^\mu \, p_1\cdot \epsilon^*(l)\,p_2\cdot \epsilon^*_V
\right. \right. 
\nonumber \\
& \hspace{1.5cm} \left. \left. -4\epsilon^{*\mu}_\bot(l)\left(  p_1\cdot \epsilon^*_V\, p_2\cdot l - p_1\cdot l \, p_2\cdot \epsilon^*_V  \right)  +4\epsilon^{*\mu}_{V\bot} \,p_2\cdot l\, p_1\cdot \epsilon^*(l)
\right]
\right\}. \label{TrD3}
\end{align}
The denominators appearing in the expression for ${\cal A}^{ab}_{\mu\bot}$ are equal to
\begin{align}
&\!\!\!(q+l)^2-m^2 = \frac{1}{2}\left[ k_\bot^2 +4m^2\left(  \frac{x}{\alpha_V} -1 \right) -\frac{x}{\alpha_V} p_{V\bot}^2 -\frac{x}{\alpha_l} l_\bot^2 \right]\;, \nonumber \\
&\!\! \!(q-xp_1)^2 -m^2= -\frac{x}{2\alpha_V}(4m^2 - p_{V\bot}^2)\;, \;\;\;  (\beta p_2 +k_\bot -q)^2 -m^2= \frac{1}{2}\!\left( k_\bot^2 -4m^2 +\frac{x}{\alpha_l} l_\bot^2  \right) . \label{CSdenom}
\end{align}
The cross section is obtained by squaring the sum of diagrams $D(i)$, i.e. by contracting this sum with its complex conjugate through the polarization tensors for the $J/\psi$ and the gluon $g(l)$ and  the projection operator related to the factorization of the gluonic PDF, namely
\begin{equation}
{\cal D}^{(1)}(J/\psi) \equiv
{\cal D}^{\mu \nu \rho} g_{\bot \mu \mu'}\left( -g_{\rho \rho'} + \frac{q_\rho q_{\rho'}}{m^2} \right) \left( -g_{\nu \nu'} +\frac{p_{2\nu}l_{\nu'} +p_{2\nu'}l_\nu }{p_2\cdot l}    \right)  {\cal D}^{ *\mu' \nu' \rho'}\,.
\end{equation}
Thus we obtain that
\begin{equation}
\sum_{\lambda_V\, \lambda_l} {\cal A}_{\mu\bot}^{ab}\, g_\bot^{\mu \nu}\, ({\cal A}_{\nu\bot}^{ab})^* =\,\frac{g^6 C_1^2}{(4N)^2}\frac{d^{abl} d^{abl}}{4^2} \,{\cal D}^{(1)}(J/\psi) \, ,
\label{CSAA*}
\end{equation}
which by taking into account eq.~(\ref{CSVertex}) gives the $J/\psi$ production vertex in the form
\begin{equation}
V_{J/\psi}^{(1)} 
= \frac{|p_{V\bot}|\sqrt{2 } g^6 C_1^2}{s^2 \pi^4 2^{13} k_\bot^2} \frac{d^{abl} d^{abl}}{N^2(N^2-1)^2} \frac{\theta(x-\alpha_V)}{x(x-\alpha_V)}{\cal D}^{(1)}(J/\psi) \, ,
\label{vertexsin}
\end{equation}
with $\alpha_V=\frac{\sqrt{4m^2-p_{V\bot}^2}}{\sqrt{s}} e^{y_V}$ and $d^{abl} d^{abl} = \frac{(N^2-4)(N^2-1)}{N}$.
The final expression for 
${\cal D}^{(1)}(J/\psi)$ reads
\begin{align}
&{\cal D}^{(1)}(J/\psi)=
\frac{2^9}{\left(m^2-q_\perp^2\right)^2 \left(4 x k_\perp\cdot
   q_\perp+k_\perp^2 \left(\alpha_V-2 x\right)+4 m^2
   \left(x-\alpha_V\right)-4 x q_\perp^2\right)^2}\nonumber \\
& \times
   \frac{s^2 \alpha_V^2 \left(\alpha_V-x\right)^2}{
   \left(4 x \left(x q_\perp^2-\alpha_V k_\perp\cdot
   q_\perp\right)+k_\perp^2 \alpha_V^2-4 m^2 \left(\alpha_V-x\right)^2\right)^2} 
 \left\{32 m^4 \alpha_V^2 \left(\alpha_V-x\right)^2 (k_\perp\cdot q_\perp)^2
 \right. \nonumber \\
& \left.
 +(k_\perp^2)^3 \alpha_V^2 \left[m^2 \left(\alpha_V^2-2 x \alpha_V+2 x^2\right)-q_\perp^2 \left(\alpha_V-x\right)^2\right] 
 \right. \nonumber \\
& \left.
+8 m^2 k_\perp^2 \left[-2 \alpha_V k_\perp\cdot q_\perp \left(m^2 \left(\alpha_V-x\right)^3+q_\perp^2 \left(2
   \alpha_V^3+2 x^2 \alpha_V-3 x \alpha_V^2+x^3\right)\right)
\right.\right.   \nonumber \\
& \left.\left.   
    +\alpha_V^2 \left(\alpha_V^2-2 x \alpha_V+3 x^2\right) (k_\perp\cdot q_\perp)^2 +2 \left(m^2 \left(\alpha_V-x\right)^2-q_\perp^2 \left(\alpha_V^2-x \alpha_V+x^2\right)\right)^2\right]
 \right.   \nonumber \\
& \left.   
       -4 (k_\perp^2)^2 \left[\alpha_V
   k_\perp\cdot q_\perp \left(m^2 \left(\alpha_V^3+x^2 \alpha_V-x \alpha_V^2+x^3\right)-x q_\perp^2 \left(\alpha_V-x\right)^2\right)+m^4 \left(\alpha_V-x\right)^4
\right.\right.   \nonumber \\
&
\left.\left.
   +m^2
   q_\perp^2 \left(-5 \alpha_V^4+6 x^3 \alpha_V-13 x^2 \alpha_V^2+12 x \alpha_V^3-2 x^4\right)+x^2 (q_\perp^2)^2 \left(\alpha_V-x\right)^2\right]\right\} \,.
\end{align}

\subsection{Color-octet NRQCD contribution}

In the color-octet contribution $[M]$ denotes one meson state, thus  
\begin{equation}
\delta(x-[\alpha_M])\delta^2(k_\bot - [p_{M\bot}]) \left[  \frac{d^3p_M}{(2\pi )^32E_M} \right] = \frac{\delta(x-\alpha_V) \delta^2(k_\bot - p_{V\bot})}{2(2\pi )^3} dy_V d^2p_{V\bot} \, ,
\label{OctetM}
\end{equation}
which leads to the differential cross section
\begin{align}
\frac{d\sigma}{dy_V\, d|p_{V\bot }|d\phi_V dy_{J} d|p_{J\bot}|d\phi_{J}} = \int & \,dx \, g(x\,) \,dy \, H^q(y) \, d^2k_\bot 
\frac{|p_{V\bot}|\delta(x-\alpha_V)  \delta^2(k_\bot - p_{V\bot})  }{\sqrt{2} \pi s^2 (N^2-1)^2 \, k_\bot^2\, x} \nonumber \\
& \times \sum_{\lambda_V}  {\cal A}_{\mu \bot}^{ab} g_\bot^{\mu \nu}({\cal A}_{\nu \bot}^{ab})^*  \,\,V_q^{(0)}(-k_\bot,y) \, ,
\label{CrSecOctet}
\end{align}
from which we read off the $J/\psi$ production vertex of the color octet NRQCD contribution:
\begin{equation}
V_{J/\psi}^{(8)}(k_\bot,x) =  \frac{|p_{V\bot}|\delta(x-\alpha_V)  \delta^2(k_\bot - p_{V\bot})  }{\sqrt{2} \pi s^2 (N^2-1)^2 \, k_\bot^2\, x} 
 \sum_{\lambda_{V}}  {\cal A}_{\mu \bot}^{ab} g_\bot^{\mu \nu}({\cal A}_{\nu \bot}^{ab})^* \, .
\label{VertexJPsiOctet }
\end{equation}
The vertex which allows to pass from open $q \bar{q}$ production to $J/\psi$ production in color octet NRQCD is defined as
\begin{equation}
[v(q)\bar u(q)]^{ij\rightarrow d}_{\alpha \beta} \rightarrow\,t^d_{ij}  d_8\,
 \left( \frac{\langle{\cal O}_8 \rangle_V}{m} \right)^{1/2}\left[ \hat \epsilon^*_V \left(  2\hat q +2m \right)\right]_{\alpha \beta} \, ,
\label{COvertex}
\end{equation}
where the  value of the coefficient $d_8$  is determined by comparison with the result of Cho and Leibovich~\cite{Cho:1995vh,Cho:1995ce}, namely eq.~(A.1b) of ref.~\cite{Cho:1995ce}, for the total 
squared amplitude for creating a specific quarkonium state $^3S_1^{(8)}$. Note that here we only consider the case where the quark-antiquark pair has the same spin and orbital momentum as the $J/\psi$ meson. At large transverse momentum, which is the case we will consider in the following, this contribution is found to be dominant, see e.g. ref.\cite{Kramer:2001hh}. For  $N=3$ the coefficient $d_8$ equals $d_8=\frac{1}{4\sqrt{3}}$.
An early analysis~\cite{Hagler:2000eu} gave for the non-perturbative coefficient
\begin{equation}
C_8 \equiv  \left( \frac{\langle{\cal O}_8 \rangle_V}{m} \right)^{1/2}
\end{equation}
values between $3.2\times 10^{-4}$ and $5 \times 10^{-4}$ GeV$^3$. More recent analyses~\cite{Butenschoen:2011yh,Chao:2012iv,Bodwin:2014gia}, as quoted in ref.~\cite{Bain:2017wvk}, obtained significantly larger values which we will use here, namely we will vary $\langle{\cal O}_8 \rangle_{J/\psi}$ between $0.224 \times 10^{-2}$ and $1.1 \times 10^{-2}$ GeV$^3$.
\def\sca{1.3}
\def\spa{1cm}
\begin{figure}[t]
\center
\psfrag{a}{}
\psfrag{b}{}
\psfrag{c}{}
\psfrag{d}{}
\psfrag{l}{$\ell$}
\psfrag{qu}{$q$}
\psfrag{qd}{$-q$}
\psfrag{p1}[R]{$xp_1$}
\psfrag{p2}[R]{$\beta p_2 + k_\perp \hspace{.6cm}$}
\begin{tabular}{ccc}
\includegraphics[scale=1.1]{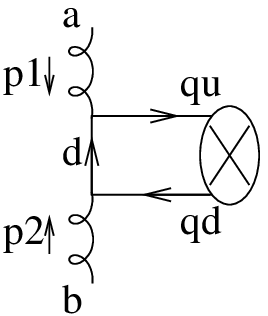}
&
\hspace{1cm}\raisebox{0cm}{\includegraphics[scale=1.1]{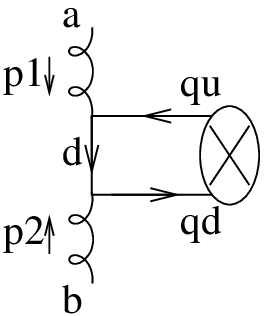}}
&
\hspace{1cm}\raisebox{0cm}{\includegraphics[scale=1.1]{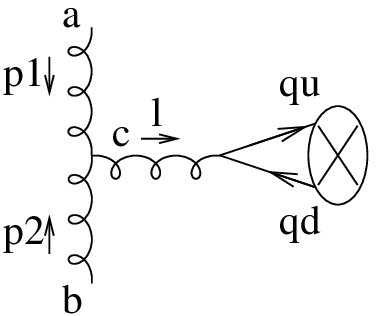}}\\
A & B & C
\end{tabular}
\caption{The 3 diagrams contributing to the amplitude in color octet NRQCD.
The blobs with a cross symbolize the Fierz structure of eq.~(\ref{COvertex}).}
\label{Fig:Diagrams-CO}
\end{figure}

The hard part corresponds to the sum of the three diagrams of figure~\ref{Fig:Diagrams-CO}, namely
\begin{align}
\label{Doctet}
{\cal A}_{ \bot}^{ab \mu}(A+B+C) = (-ig)^2 i \,\bar u(q) & \left[ t^at^b \, \gamma_\bot^\mu \frac{\hat q - x \hat p_1 +m}{(q-x p_1)^2-m^2} \hat p_2  + t^bt^a \hat p_2 \frac{x \hat p_1 - \hat q +m}{(x p_1-q)^2-m^2} \gamma^\mu_\bot \right. \nonumber \\
& \hspace{0.2cm} \left. -if^{abc} \left(   -2k_\bot^\mu p_2^\rho +4 p_2\cdot q g_\bot^{\mu \rho}   \right) \frac{t^c \gamma_\rho}{4q^2}
\right] v(q) \,.
\end{align}
After taking into account the projection (\ref{COvertex}) we obtain
\begin{align}
\label{DOctetJ}
{\cal A}_{ \bot}^{ab \mu}(A+B+C \rightarrow J/\psi)_8 = (-ig)^2i^2 f^{abd} \,\frac{1}{2} & \left\{ \frac{8m}{(q-xp_1)^2-m^2} \left[     -2q\cdot p_2
\epsilon^{*\mu}_{V\bot} + k_\bot^\mu p_2\cdot \epsilon^*_V \right]  \right.
\nonumber \\
& \hspace{0.2cm} \left. -\frac{16m}{4q^2} \left[   - k_\bot^\mu p_2\cdot \epsilon^*_\bot + 2p_2\cdot q  \epsilon^{*\mu}_{V\bot}   \right]
 \right\}d_8\,{\cal C}_8\, ,
\end{align}
in which the propagator $(q-x p_1)^2-m^2=-\frac{1}{2}(4m^2- k_\bot^2)$. 
One can easily check that this sum vanishes in the limit $k_\perp \to 0$, as it should be the case for an impact factor in $k_t$-factorization. This is also true at the level of open quark production, see eq.~(\ref{Doctet}).
The result (\ref{DOctetJ}) together with eq.~(\ref{VertexJPsiOctet }) leads to the $J/\psi$ production vertex for $N=3$:
\begin{equation}
V_{J/\psi}^{(8)} =
 -\delta (x-\alpha_V) \delta^2(k_\bot - p_{V \bot}) \frac{|p_{V \bot}|\sqrt{2} g^4 k_\bot^2 x }{128\pi m^3(4m^2-k_\bot^2)^2} \langle{\cal O}_8 \rangle_V \,.
\label{Voctet}
\end{equation}

\subsection{Color evaporation model}
In the color evaporation model $[M]$ denotes an open quark-antiquark produced state with an invariant mass $M$. Moreover, the differential cross section in this model involves  an integration over the invariant mass $M^2$ in the interval $[4m^2,4M_D^2]$, as it is assumed that in this interval below the $D-$meson mass threshold, a fixed fraction of these $c \bar{c}$ pairs (either produced in a singlet or in an octet color state) will form $J/\psi$ bound states. This fraction is parametrized by the constant $F_{J/\psi}$, which is assumed to
be universal as one of the main assumptions of the color evaporation model, and we will vary it between 0.02 and 0.04 based on a recent  analysis~\cite{Nelson:2012bc}.

The $J/\psi$ momentum in this model is the sum $k_J=k_1+k_2$. We parametrize the momentum $k_1$ of the produced quark and the momentum $k_2$ of the produced anti-quark as follows:
\begin{align}
k_1 &= \alpha_1 p_1 + \beta_1p_2 +k_{1\bot} \equiv x \alpha p_1 + \frac{m^2- (\alpha k_\bot +l_\bot )^2}{x \alpha s} p_2 + \alpha k_\bot +l_\bot\,, \;\;\;k_1^2=m^2, \label{CEkin1} \\
k_2 &= \alpha_2 p_1 + \beta_2p_2 +k_{2\bot} \equiv x \bar \alpha p_1 + \frac{m^2- (\bar \alpha k_\bot - l_\bot)^2}{x\bar \alpha s}p_2 + \bar \alpha k_\bot - l_\bot\,, \;\;\;k_2^2=m^2, \label{CEkin2} \\
M^2 &= (k_1 + k_2)^2 \equiv \frac{m^2-l_\bot^2}{\alpha \bar \alpha}\;,
\label{CEkinM}
\end{align}
with $\bar\alpha=1-\alpha\,.$
Thus,
\begin{align}
\delta(x\!-\![\alpha_M])\delta^2(k_\bot\!-\![p_{M\bot}]) 
\!\left[\!\frac{d^2p_M}{(2\pi )^32E_M}\!\right] \!\!&=\!
\delta(x\!-\!\alpha_1\!-\!\alpha_2)\delta^2(k_\bot\!-\!k_{1\bot}\!-\!k_{2\bot}) \frac{d^3k_1}{(2\pi )^32E_1}  \frac{d^3k_2}{(2\pi )^32E_2} \nonumber \\
&=\delta(x-\alpha_V) \delta^2(k_\bot - k_{V\bot}) \frac{1}{4(2\pi )^6}\, \frac{d\alpha\, d^2l_\bot}{\alpha \bar \alpha} \,dy_V d^2k_{V\bot}\,,
\label{CEPhSp}
\end{align}
which leads, by taking into account (\ref{CSecM}), to the differential cross section in the color evaporation model having the form
\begin{align}
&\frac{d\sigma}{dy_V d|p_{V\bot}|d\phi_V dy_{J}d|p_{J\bot}|d\phi_{J}}= F_{J/\psi} \!\int \! dx \, g(x) dy H^q(dy) d^2k_\bot  \! \! \int_{4m^2}^{4M^2_{D}} \!\! dM^2\,
\delta \left(\! M^2 - \frac{m^2 - l^2_\bot }{\alpha \bar \alpha}\right)\!
\nonumber \\
& \hspace{0.6cm}\times \frac{ |p_{V \perp}|\sqrt{2} \,\delta(x-\alpha_V) \delta^2(k_\bot - p_{V\bot})}{2^5 \pi^4 s^2 (N^2-1)^2\, k^2_\bot\, x}\frac{d\alpha\, d^2l_\bot}{\alpha \bar \alpha}  \sum_{\lambda_{k_1}\lambda_{k_2}}  {\cal A}_{i \bot}^{ab} g_\bot^{i j}({\cal A}_{j \bot}^{ab})^*  \,\,V_q^{(0)}(k_\bot,y)\,,
\label{CECrSec}
\end{align}
with $y_V= \ln \left( \alpha_V / \sqrt{\frac{M^2-p^2_{V\bot}}{s}} \right)$, from which we read off the $J/\psi$ production vertex in the color evaporation model:
\begin{align}
V^{(\rm CEM)}_{J/\psi}(k_\bot,x) = F_{J/\psi}\,\int\limits_{4m^2}^{4M_{D}^2} & dM^2 \, \delta\left(M^2- \frac{m^2 -l_\bot^2}{\alpha \bar \alpha}\right)\frac{d\alpha\, d^2l_\bot}{\alpha \bar \alpha} \sum_{\lambda_{k_1}\lambda_{k_2}}  {\cal A}_{i \bot}^{ab} g_\bot^{i j}({\cal A}_{j \bot}^{ab})^* \nonumber \\
& \times \frac{|p_{V\bot}|\sqrt{2} \,\delta\left(x-\alpha_V) \delta^2(k_\bot - p_{V\bot}\right)}{2^5 \pi^4 s^2 (N^2-1)^2\, k^2_\bot\, x}  \,. \label{CEVertex}
\end{align}

\begin{figure}[t]
\center
\psfrag{a}{}
\psfrag{b}{}
\psfrag{c}{}
\psfrag{d}{}
\psfrag{l}{$\ell$}
\psfrag{qu}{$q$}
\psfrag{qd}{$-q$}
\psfrag{p1}[R]{$xp_1$}
\psfrag{p2}[R]{$\beta p_2 + k_\perp \hspace{.6cm}$}
\begin{tabular}{ccc}
\includegraphics[scale=1.1]{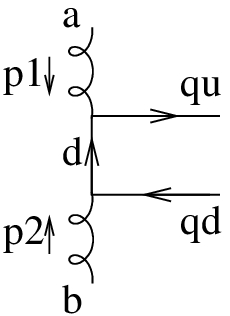}
&
\hspace{1cm}\raisebox{0cm}{\includegraphics[scale=1.1]{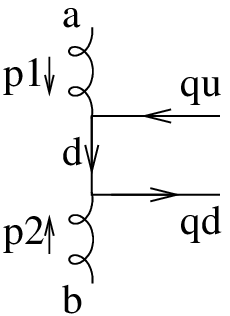}}
&
\hspace{1cm}\raisebox{0cm}{\includegraphics[scale=1.1]{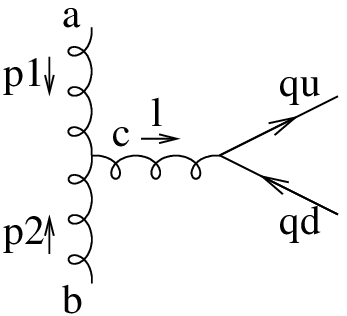}}
\end{tabular}
\caption{The 3 diagrams contributing to the amplitude in the color evaporation model.}
\label{Fig:Diagrams-Cevaporation}
\end{figure}

The contribution to the hard part of the vertex in the Born approximation is given by three diagrams analogous to the ones of the color octet NRQCD contribution, except for the absence of any Fierz projection, since we simply deal with open quark-antiquark production.
These diagrams are shown in figure~\ref{Fig:Diagrams-Cevaporation}.
The hard part then reads 
\def\sca{1.3}
\def\spa{1cm}
\begin{align}
{\cal A}_{i\bot}^{ab} \!=\! \bar u(k_1) & \! \left[ \! (-ig\gamma_{i\bot} t^a) \frac{i(-x\hat p_1 \! + \! \hat k_1 \!+ \!m)}{(-xp_1\!+\!k_1)^2\! -\! m^2} (-ig t^b \hat p_2) + (-ig t^b \hat p_2) \frac{i(x \hat p_1 \! - \! \hat k_2 \! + \! m)}{(xp_1 \! - \! k_2)^2 \! - \! m^2}(-ig \gamma_{i\bot} t^a) \right. \nonumber \\
& \hspace{0.2cm} \left. +g f^{abc} \left(  -2 p_2^\nu k_{i\bot} + g^{ \nu}_{i\bot} xs  \right) \frac{(-i)}{M^2} (-ig\gamma_\nu t^c)
\right]   v(k_2)\;. \label{Aiab}
\end{align}
Thus its contribution to the $J/\psi$ production vertex has the form
\begin{equation}
\sum_{\lambda_{k_1}\lambda_{k_2}}     ({\cal A}^{ab}_{i\bot})^*   g_{\bot }^{ij}   {\cal A}^{ab}_{j\bot} = \frac{g^4}{4} \left(   c_a \Tr_a + c_b \Tr_b  \right) \, ,
\label{CEhard}
\end{equation}
where the two color structures are given by
\begin{equation}
c_a=\frac{f^{abc}f^{abc}}{2}=\frac{N(N^2-1)}{2}\,,\, c_b= \frac{\delta^{ab}\delta^{ab}}{N^2}+\frac{d^{abc}d^{abc}}{2}=\frac{N^2-1}{N^2}\left( 1 + \frac{N(N^2-4)}{2}\right),\,\,\,\,\,\,\,
\label{ColorStructure-evap}
\end{equation}
and the two corresponding coefficients read
\begin{align}
\Tr_a=&-4s \left[ 
\alpha_2 \beta_1 \left(\! -\frac{1}{\beta_1} +\frac{2xs}{M^2} \right)^2
\!\!+
\alpha_1 \beta_2 \left(\! -\frac{1}{\beta_2} +\frac{2xs}{M^2} \right)^2
\right] -8m^2 \left(  \frac{2xs}{M^2}  - \frac{1}{\beta_2}   \right) \! \left(  \frac{2xs}{M^2}  - \frac{1}{\beta_1}   \right) \nonumber \\
&+ \frac{8}{x} \left[\frac{k_{1\bot}}{\beta_1}+\frac{k_{2\bot}}{\beta_2}
- \frac{2 x s}{M^2} k_{\bot} \right] \cdot
\left[ \alpha_2 \left(\left(\frac{\alpha_1}x -1 \right)\frac{1}\beta_1 
+ \frac{2 x s}{M^2} \right) k_{1\bot} \right. \nonumber \\
& \hspace{5cm} \left. + \alpha_1 \left(\left(\frac{\alpha_2}x -1 \right)\frac{1}\beta_2 
+ \frac{2 x s}{M^2} \right) k_{2\bot}
-  \frac{ 2 \alpha_1 \alpha_2s}{M^2}  k_{\bot}\right]
\nonumber \\
=&-4s x\left[ 
\bar{\alpha} \beta_1 \left(\! -\frac{1}{\beta_1} +\frac{2xs}{M^2} \right)^2
\!\! +
\alpha \beta_2 \left(\! -\frac{1}{\beta_2} +\frac{2xs}{M^2} \right)^2
\right]
-8m^2 \left(  \frac{2xs}{M^2}  - \frac{1}{\beta_2}   \right) \left(  \frac{2xs}{M^2}  - \frac{1}{\beta_1} \right) \nonumber \\
&+\!8\!\left[\frac{k_{1\bot}}{\beta_1}+\frac{k_{2\bot}}{\beta_2}
- \frac{2 x s}{M^2} k_{\bot} \right] \! \cdot \!
\left[ \bar{\alpha} \! \left(\!-\frac{\bar{\alpha}}\beta_1 
\!+\! \frac{2 x s}{M^2} \right) k_{1\bot}
\!+\! \alpha \! \left(\!-\frac{\alpha}\beta_2 
+ \frac{2 x s}{M^2} \right) k_{2\bot}
-  \alpha \bar{\alpha} \frac{2 x s}{M^2}  k_{\bot}\right] \! , \label{Tra}
\end{align}
and
\begin{align}
\Tr_b \!=&\! 
-4 s \left(\frac{\alpha_2}{\beta_1}+\frac{\alpha_1}{\beta_2}\right) 
+ 8 \left[ \frac{\alpha_2}x \left(\frac{\alpha_1}x -1 \right) \frac{k_{1\bot}}{\beta_1} 
-  \frac{\alpha_1}x \left(\frac{\alpha_2}x -1 \right) \frac{k_{2\bot}}{\beta_2} \right] \! \cdot \! \left[\frac{k_{1\bot}}{\beta_1} - \frac{k_{2\bot}}{\beta_2}   \right] + \frac{8m^2}{\beta_1 \beta_2} \,\,\,
\nonumber \\
\!=&\!
-4 s x\left(\frac{\bar{\alpha}}{\beta_1}+\frac{\alpha}{\beta_2}\right) 
- 8 \left[ \bar{\alpha}^2 \frac{k_{1\bot}}{\beta_1} 
-  
\alpha^2
\frac{k_{2\bot}}{\beta_2} \right] \! \cdot \! \left[\frac{k_{1\bot}}{\beta_1} - \frac{k_{2\bot}}{\beta_2}   \right] + \frac{8m^2}{\beta_1 \beta_2} \, . \label{Trb}
\end{align}
Using the fact that 
\begin{equation}
k_{1\perp} \underset{k_\perp \to 0}{\sim} l_\perp \, , 
\quad k_{2\perp} \underset{k_\perp \to 0}{\sim} -l_\perp \, ,
\quad
\beta_1 \underset{k_\perp \to 0}{\sim} \frac{m^2 -l_\perp^2}{x \alpha s} \, ,
\quad
\beta_2 \underset{k_\perp \to 0}{\sim} \frac{m^2 -l_\perp^2}{x \bar{\alpha} s} \, ,
\end{equation}
as well as the kinematical relation (\ref{CEkinM}), one can easily check that, as expected, both  $\Tr_a$ and $\Tr_b$  vanish in the limit $k_\perp \to 0\,.$

\section{Results}

In this section we compare the cross sections and azimuthal correlations between the $J/\psi$ meson and the jet obtained with the color singlet, color octet and color evaporation hadronization mechanisms, for two different values of the center of mass energy: $\sqrt{s}=8$ TeV and $\sqrt{s}=13$ TeV. We consider equal values of the transverse momenta of the $J/\psi$ and the jet, $|p_{V \bot}|=|p_{J \bot}|=p_\bot$, and four different kinematical configurations:
\begin{itemize}
\item $0<y_V<2.5, \; -6.5<y_J<5, \; p_\bot=10$ GeV,
\item $0<y_V<2.5, \; -4.5<y_J<0, \; p_\bot=10$ GeV,
\item $0<y_V<2.5, \; -4.5<y_J<0, \; p_\bot=20$ GeV,
\item $0<y_V<2.5, \; -4.5<y_J<0, \; p_\bot=30$ GeV.
\end{itemize}
The very backward jet in the first configuration could be measured for example with the CASTOR detector at CMS. An experimental study combining the CASTOR detector to tag the jet and the CMS tracking system to measure the $J/\psi$ meson would therefore allow to probe rapidity separations $Y \equiv y_V-y_J$ up to values as large as 9. For the other three configurations we restrict the rapidity of the jet to $y_J>-4.5$ which corresponds to the typical values accessible by the main detectors at ATLAS and CMS. In this case the maximum rapidity separation is $Y=7$. Since a BFKL calculation is valid only for a large rapidity separation, we will only show results for $Y>4$. We use the BLM renormalization scale fixing procedure, see ref.~\cite{Ducloue:2013bva}, which modifies the ``natural'' initial scale
$\mu_{R, {\rm init}}=\sqrt{|p_{V \bot}|\cdot|p_{J \bot}|}$
by
\begin{equation}
  \mu^2_{R,{\rm BLM}}= |p_{V \bot}|\cdot|p_{J \bot}| \exp \left[ \frac{1}{2}
  \chi_0(n,\gamma)-\frac{5}{3}+2\left(\!1+\frac{2}{3}I\!\right)\! \right] \, ,
\end{equation}
where
\begin{equation}
  \chi_0(n,\gamma) = 2\psi(1)-\psi\left(\gamma+\frac{n}
  {2}\right)-\psi\left(1-\gamma+\frac{n}{2}\right)\,,
\end{equation}
is the LL BFKL eigenvalue and 
$I=-2\int_0^1 dx \ln(x)/[x^2-x+1] \simeq 2.3439$. The uncertainty band is computed in the same way as in ref.~\cite{Ducloue:2013bva} with the addition of the variation of the non-perturbative constants related to $J/\psi$ hadronization in the ranges specified in the previous sections. We fix the charm quark mass to $m=1.5$ GeV.

\def\figscale{0.75}

\psfrag{sigma}{\raisebox{1.5mm}{\scalebox{0.8}{$\displaystyle \frac{d\sigma}{d|p_{V \bot}|\, d|p_{J \bot}| \, dY} [{\rm nb.GeV}^{-2}]$}}}
\psfrag{cos}{\scalebox{0.9}{$\langle \cos \varphi \rangle$}}
\psfrag{Y}{\scalebox{0.9}{$Y$}}

\psfrag{2jets}{\scalebox{0.8}{2 jets}}
\psfrag{CS}{\scalebox{0.8}{Color singlet}}
\psfrag{CO}{\scalebox{0.8}{Color octet}}
\psfrag{CEM}{\scalebox{0.8}{Color evaporation}}

\begin{figure}[t]
\hspace{-0.3cm}\includegraphics[scale=\figscale]{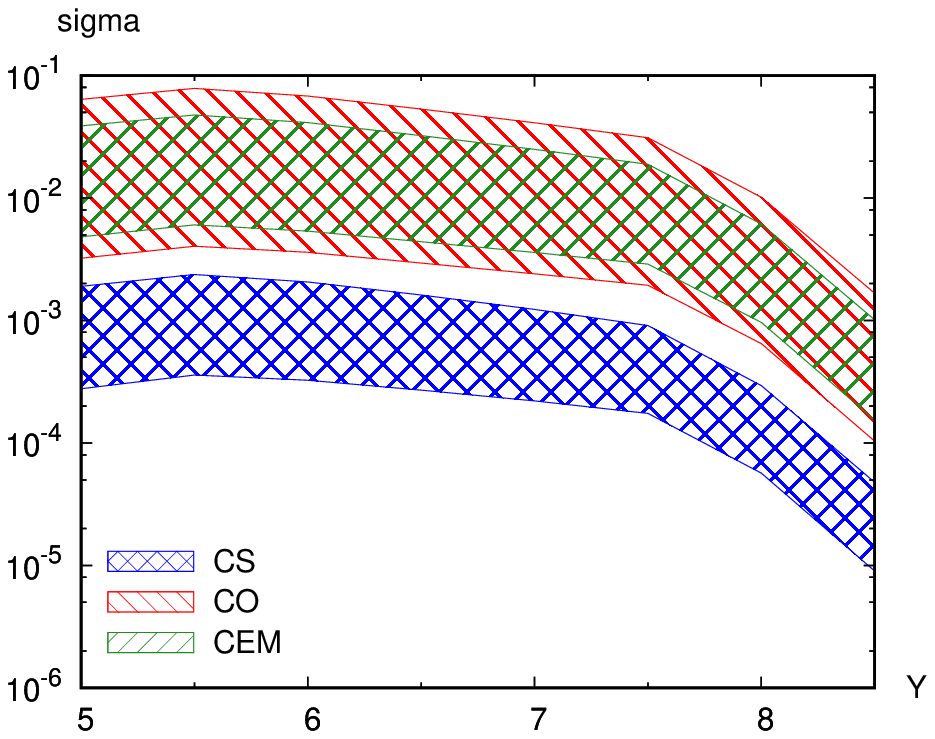}
\includegraphics[scale=\figscale]{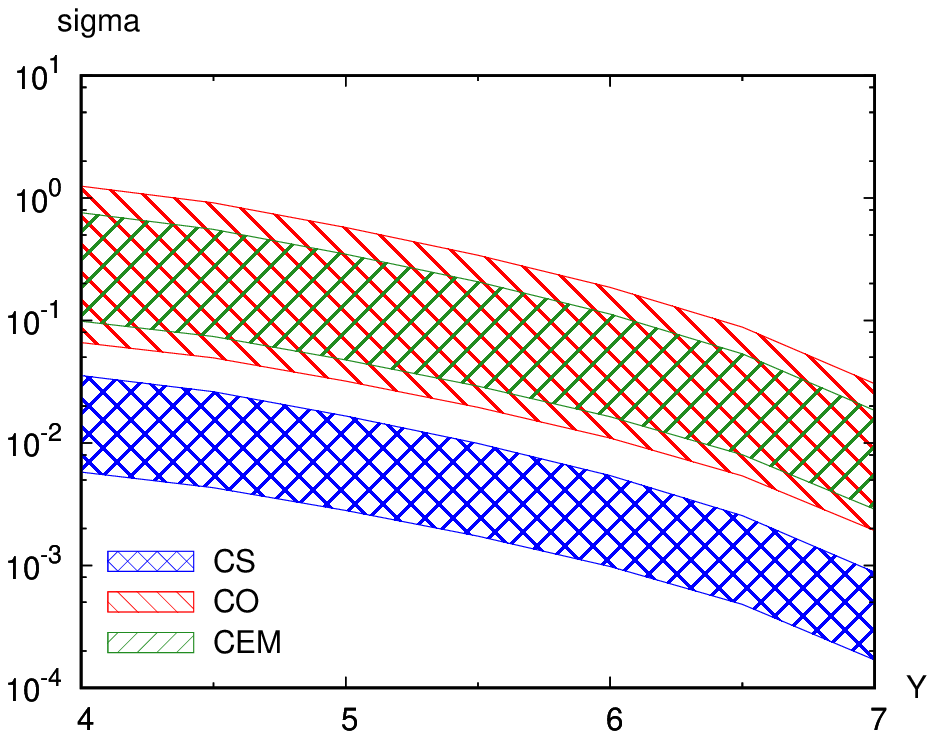}
\hspace*{0.1cm} {\small $0<y_V<2.5, \; -6.5<y_J<-5, \; p_\bot=10$ GeV} \hspace{0.8cm} {\small $0<y_V<2.5, \; -4.5<y_J<0, \; p_\bot=10$ GeV}

\vspace{0.4cm}

\hspace{-0.3cm}\includegraphics[scale=\figscale]{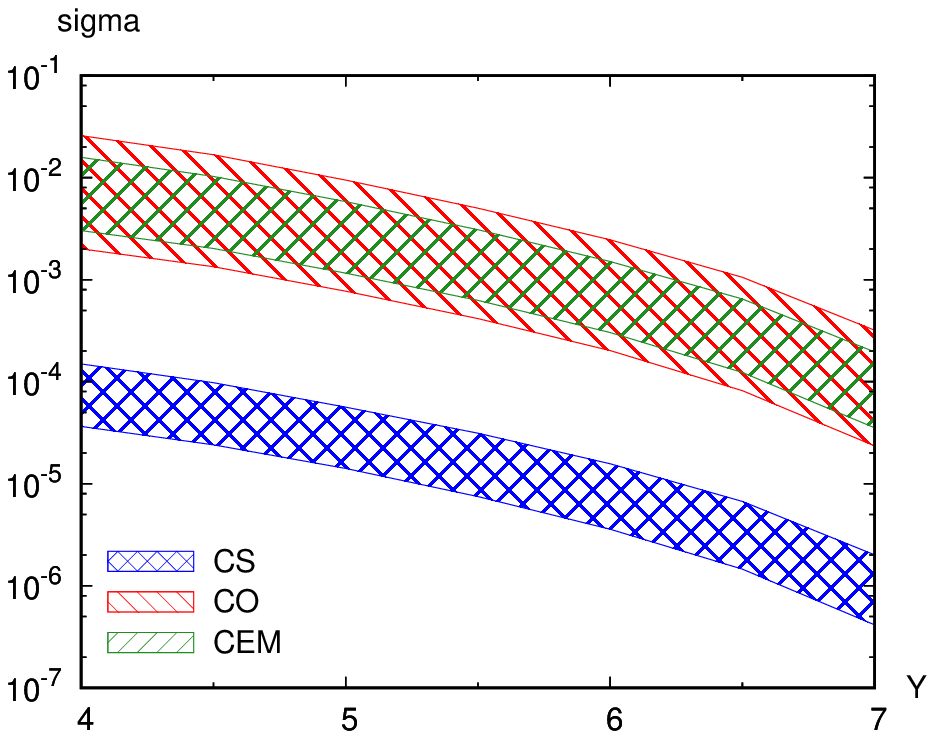}
\includegraphics[scale=\figscale]{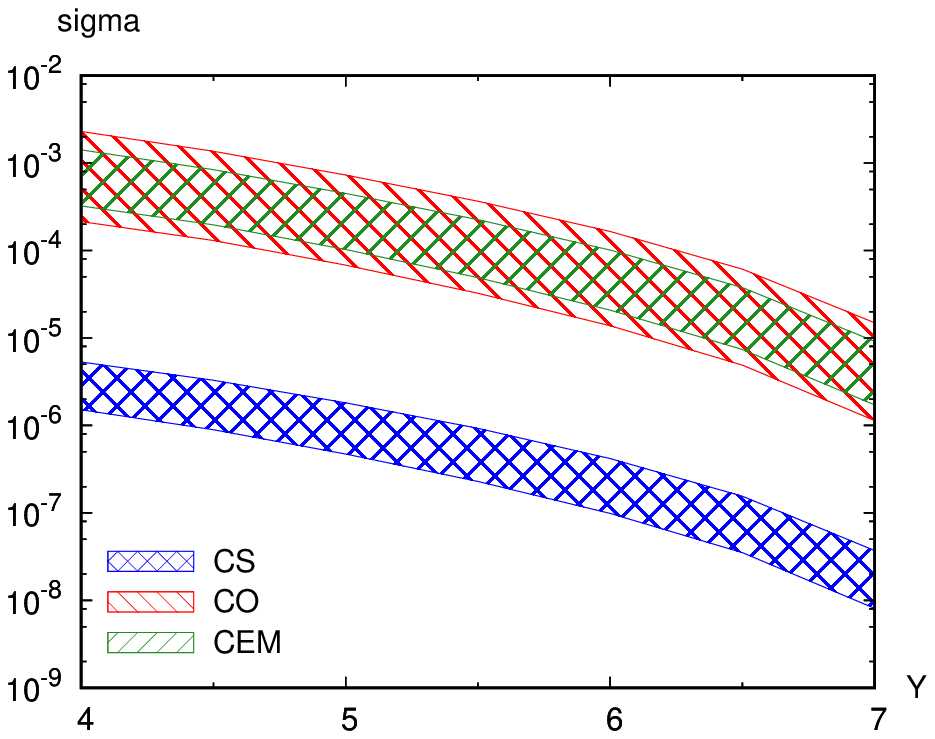}
\hspace*{0.1cm} {\small $0<y_V<2.5, \; -4.5<y_J<0, \; p_\bot=20$ GeV} \hspace{0.8cm} {\small $0<y_V<2.5, \; -4.5<y_J<0, \; p_\bot=30$ GeV}
\caption{Cross section at $\sqrt{s}=8$ TeV as a function of the relative rapidity $Y$ between the $J/\psi$ and the jet, in four different kinematical configurations.}
\label{Fig:sigma-8}
\end{figure}
\begin{figure}[t]
\hspace{-0.3cm}\includegraphics[scale=\figscale]{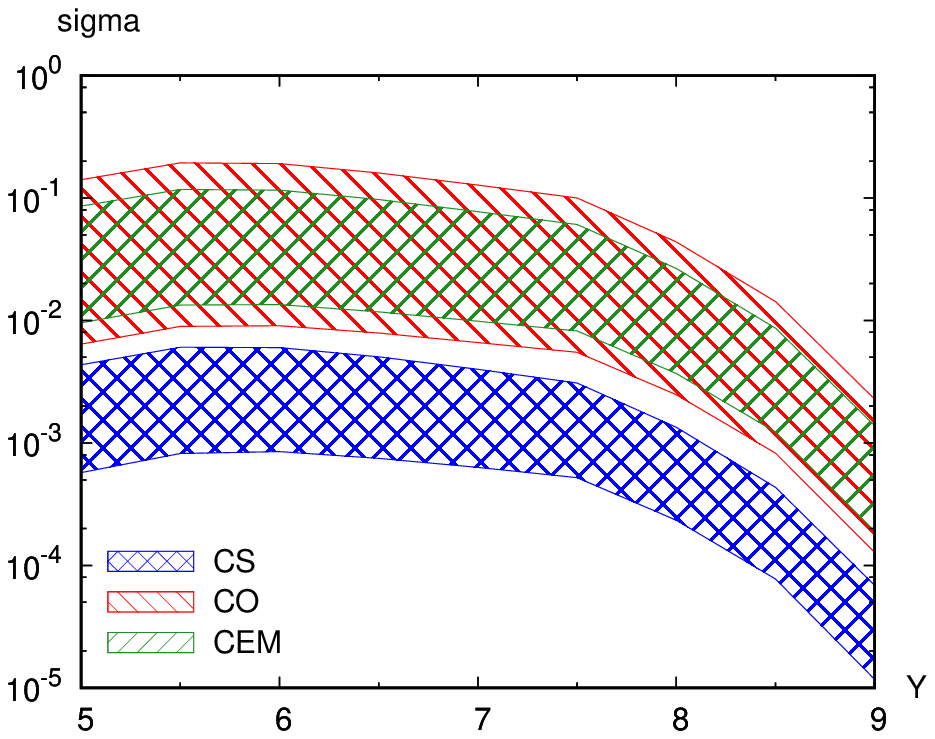}
\includegraphics[scale=\figscale]{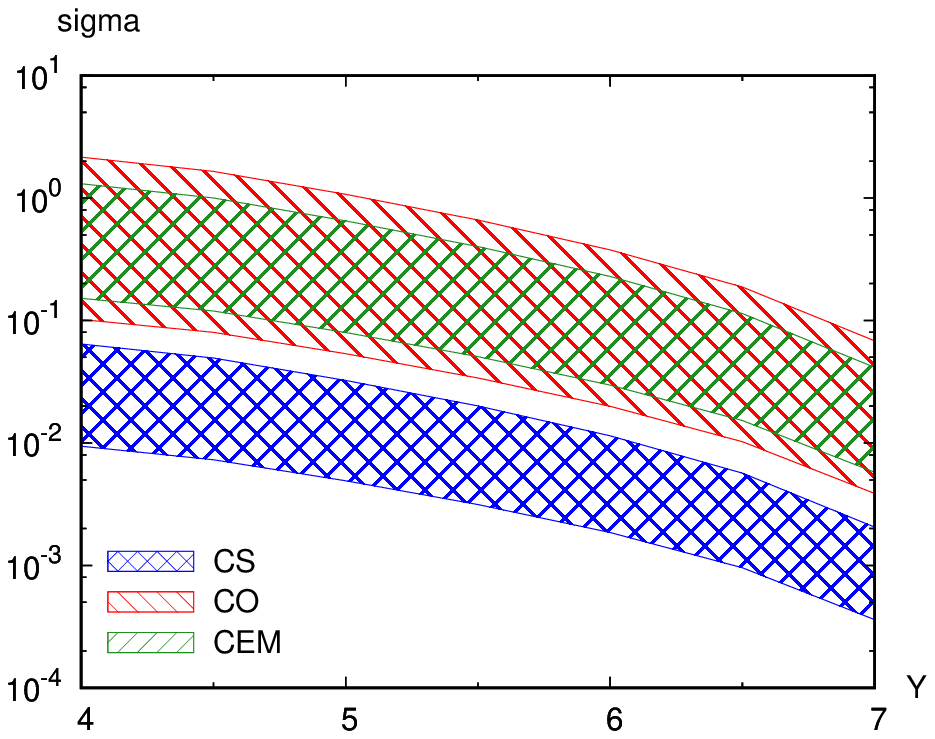}
\hspace*{0.1cm} {\small $0<y_V<2.5, \; -6.5<y_J<-5, \; p_\bot=10$ GeV} \hspace{0.8cm} {\small $0<y_V<2.5, \; -4.5<y_J<0, \; p_\bot=10$ GeV}

\vspace{0.4cm}

\hspace{-0.3cm}\includegraphics[scale=\figscale]{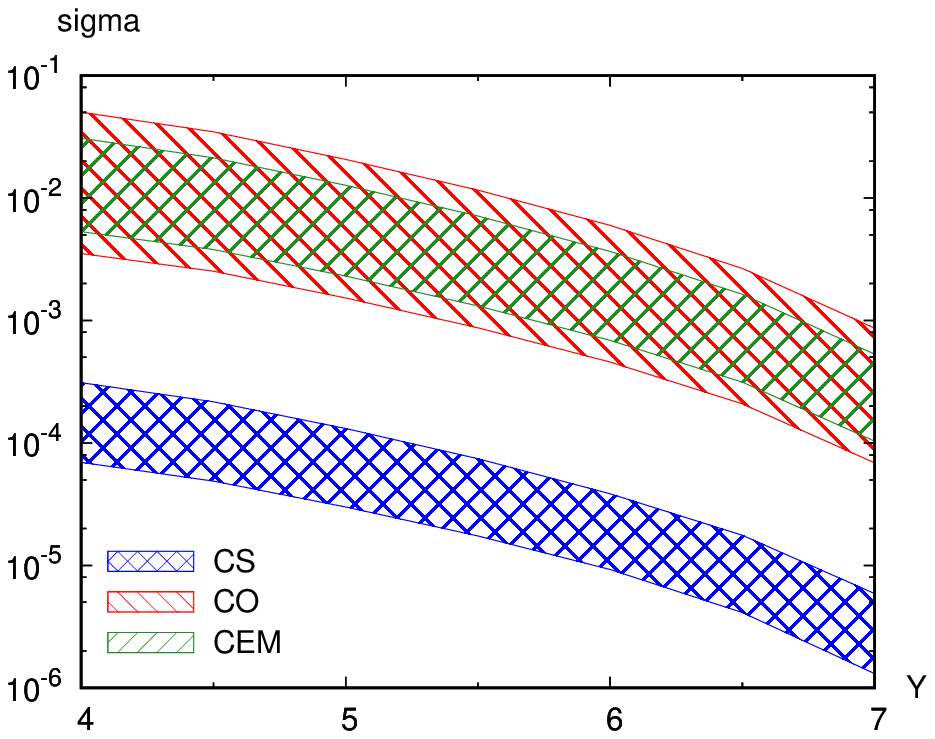}
\includegraphics[scale=\figscale]{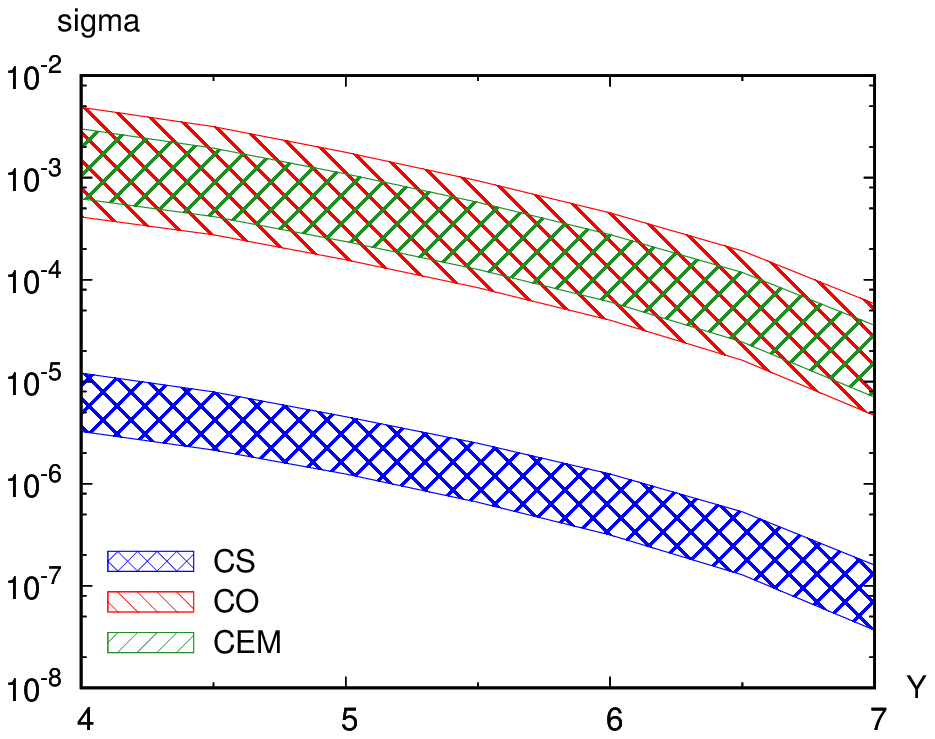}
\hspace*{0.1cm} {\small $0<y_V<2.5, \; -4.5<y_J<0, \; p_\bot=20$ GeV} \hspace{0.8cm} {\small $0<y_V<2.5, \; -4.5<y_J<0, \; p_\bot=30$ GeV}
\caption{Cross section at $\sqrt{s}=13$ TeV as a function of the relative rapidity $Y$ between the $J/\psi$ and the jet, in four different kinematical configurations.}
\label{Fig:sigma-13}
\end{figure}

In figures~\ref{Fig:sigma-8} and \ref{Fig:sigma-13}
we show the differential cross section $\frac{d\sigma}{d|p_{V \bot}|\, d|p_{J \bot}| \, dY}$ as a function of the rapidity separation $Y$ for the four kinematical cuts described above, for $\sqrt{s}=8$ TeV and $\sqrt{s}=13$ TeV respectively.
We observe that in NRQCD the color octet contribution dominates over the color singlet one, especially at high $p_\bot$. The color evaporation model leads to similar results as the color octet NRQCD contribution. Note, however, that the absolute normalization of the cross section in the color evaporation model is not very well determined.
As expected, the cross-sections slightly increase when passing from 
$\sqrt{s}=8$ TeV to $\sqrt{s}=13$ TeV, although this increase is much smaller than the uncertainties.

\begin{figure}[t]
\hspace{-0.3cm}\includegraphics[scale=\figscale]{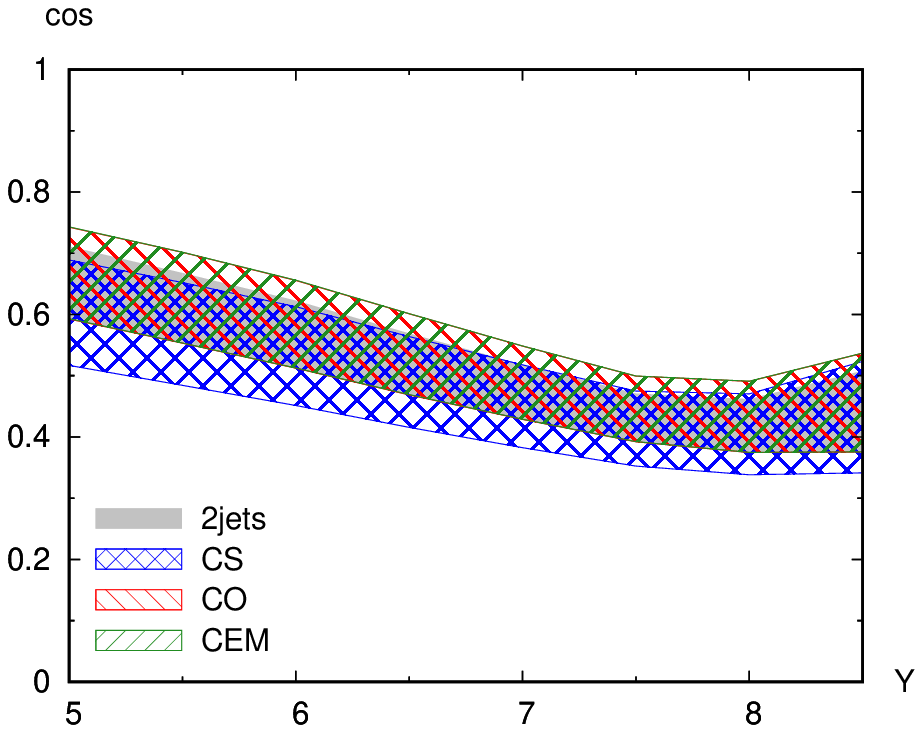}
\includegraphics[scale=\figscale]{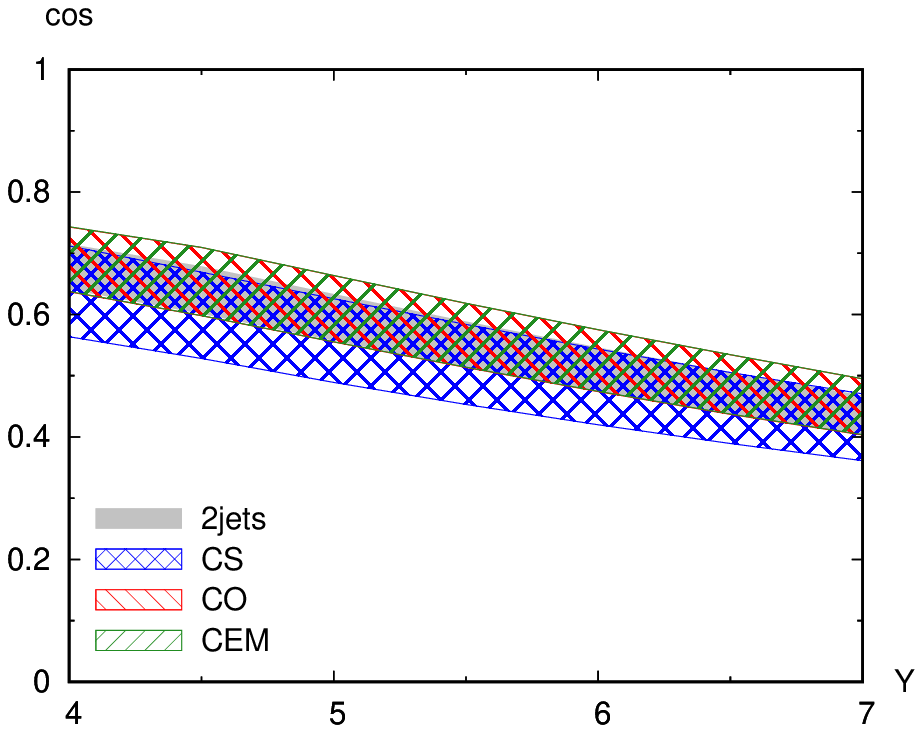}
\hspace*{0.1cm} {\small $0<y_V<2.5, \; -6.5<y_J<-5, \; p_\bot=10$ GeV} \hspace{0.8cm} {\small $0<y_V<2.5, \; -4.5<y_J<0, \; p_\bot=10$ GeV}

\vspace{0.1cm}

\hspace{-0.3cm}\includegraphics[scale=\figscale]{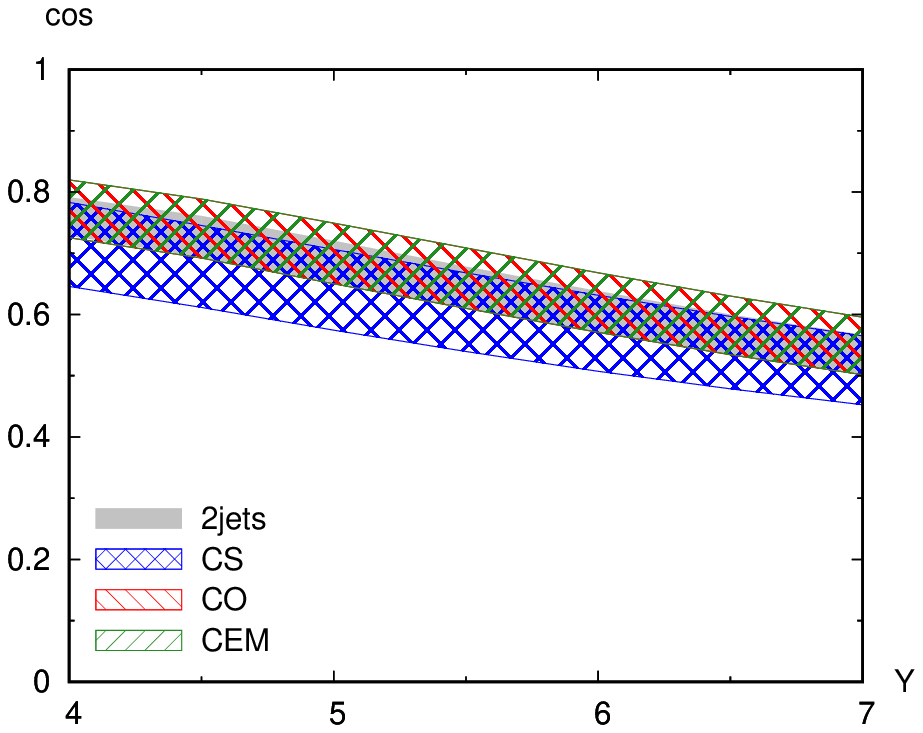}
\includegraphics[scale=\figscale]{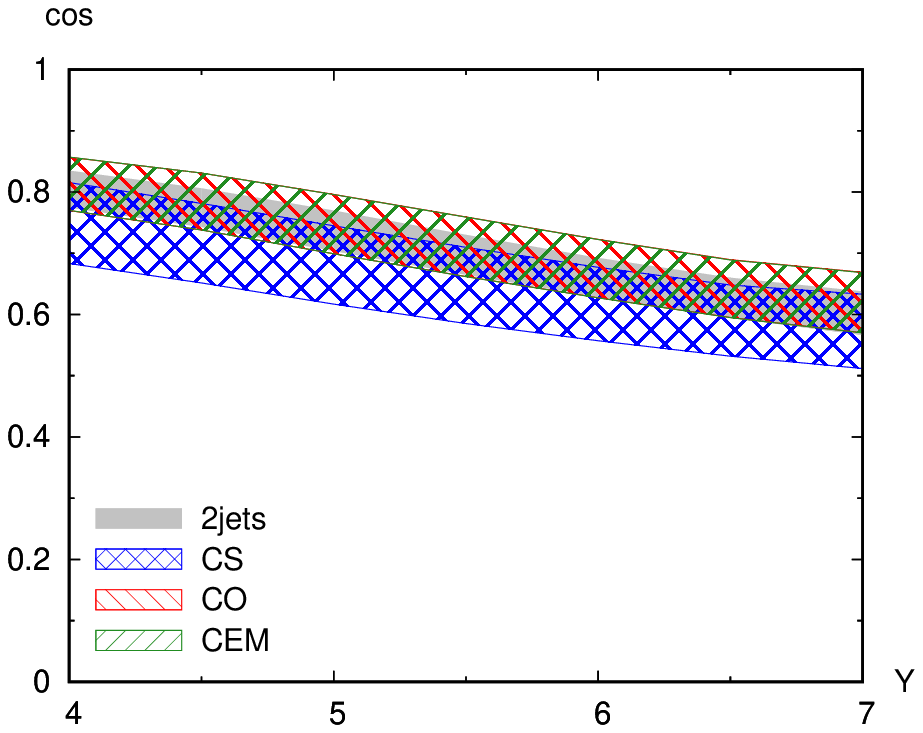}
\hspace*{0.1cm} {\small $0<y_V<2.5, \; -4.5<y_J<0, \; p_\bot=20$ GeV} \hspace{0.8cm} {\small $0<y_V<2.5, \; -4.5<y_J<0, \; p_\bot=30$ GeV}
\caption{Variation of $\langle \cos \varphi \rangle$ at $\sqrt{s}=8$ TeV as a function of the relative rapidity $Y$ between the $J/\psi$ and the jet, for the four kinematical cuts described in the text. The grey band corresponds to the results obtained when the $J/\psi$ production vertex is replaced by the leading order jet production vertex.}
\label{Fig:cos-8}
\end{figure}

\begin{figure}[t]
\hspace{-0.3cm}\includegraphics[scale=\figscale]{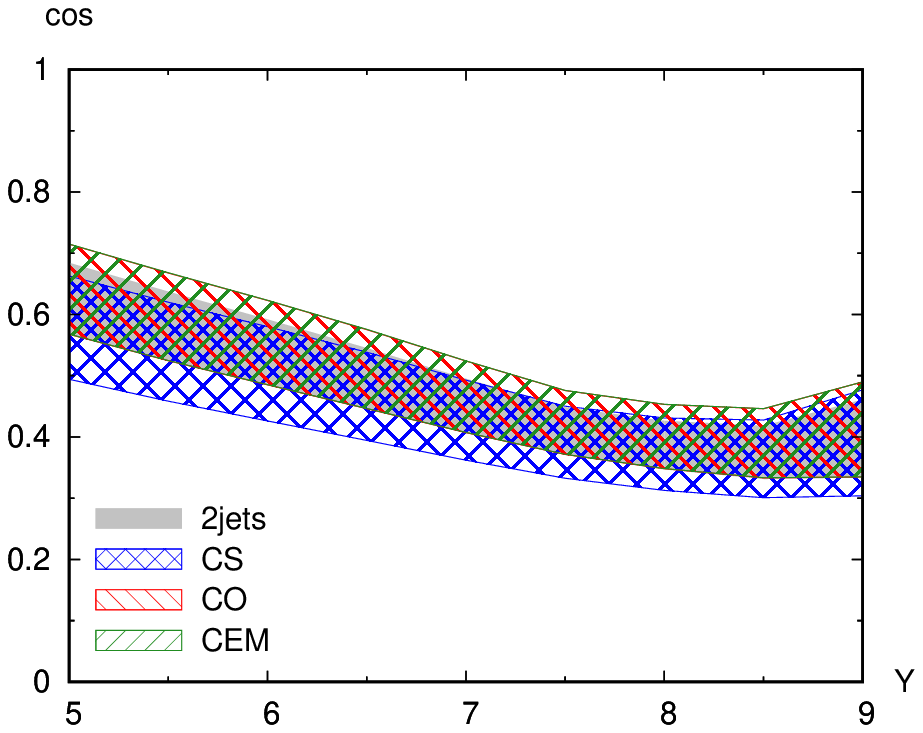}
\includegraphics[scale=\figscale]{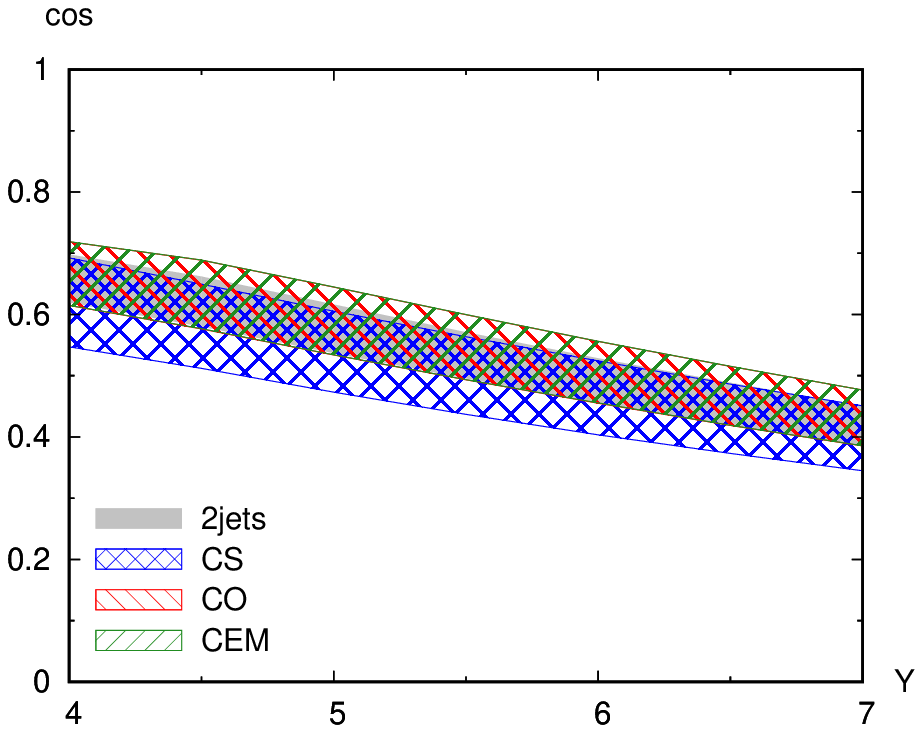}
\hspace*{0.1cm} {\small $0<y_V<2.5, \; -6.5<y_J<-5, \; p_\bot=10$ GeV} \hspace{0.8cm} {\small $0<y_V<2.5, \; -4.5<y_J<0, \; p_\bot=10$ GeV}

\vspace{0.1cm}

\hspace{-0.3cm}\includegraphics[scale=\figscale]{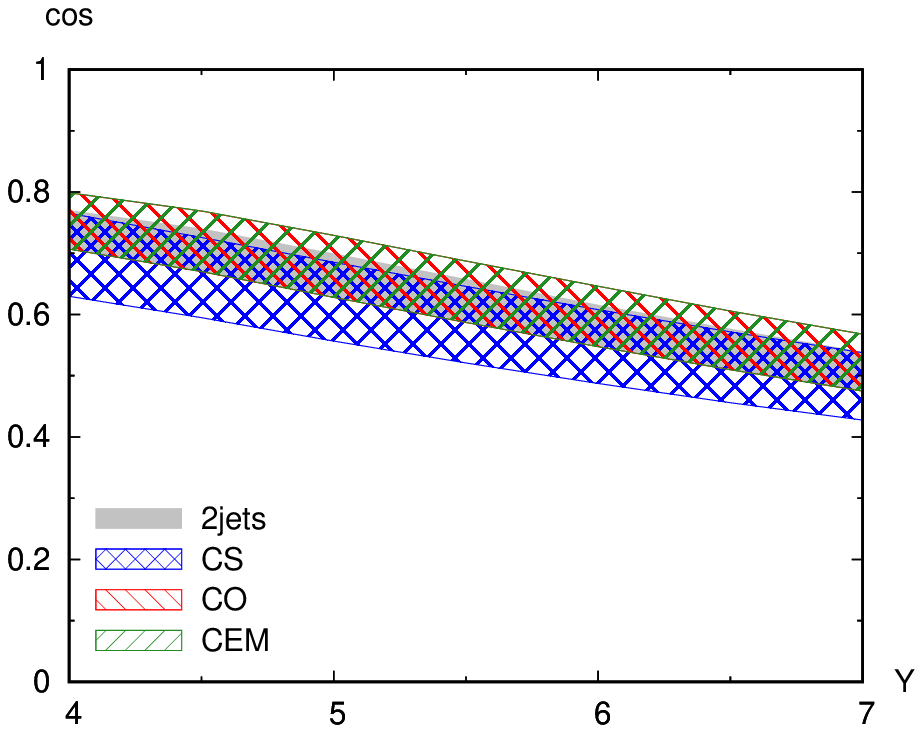}
\includegraphics[scale=\figscale]{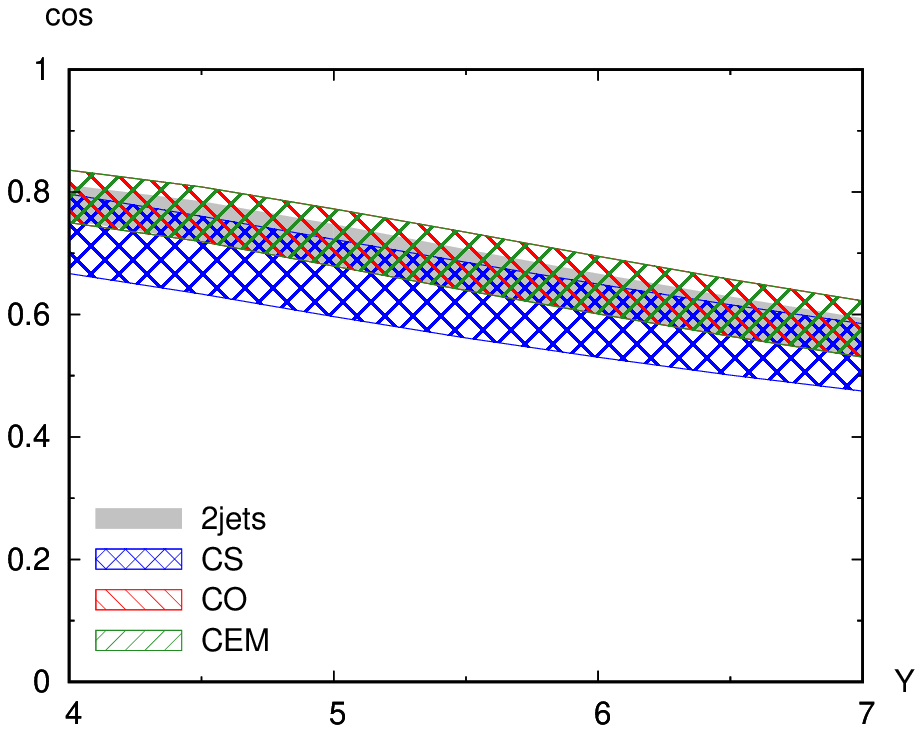}
\hspace*{0.1cm} {\small $0<y_V<2.5, \; -4.5<y_J<0, \; p_\bot=20$ GeV} \hspace{0.8cm} {\small $0<y_V<2.5, \; -4.5<y_J<0, \; p_\bot=30$ GeV}
\caption{Variation of $\langle \cos \varphi \rangle$ at $\sqrt{s}=13$ TeV as a function of the relative rapidity $Y$ between the $J/\psi$ and the jet, for the four kinematical cuts described in the text. The grey band corresponds to the results obtained when the $J/\psi$ production vertex is replaced by the leading order jet production vertex.}
\label{Fig:cos-13}
\end{figure}

In figures~\ref{Fig:cos-8} and \ref{Fig:cos-13} we show, in the same kinematics, the variation of $\langle \cos \varphi \rangle$ as a function of $Y$, where $\varphi$ is defined as $\varphi=|\phi_V-\phi_J-\pi|$, for $\sqrt{s}=8$ TeV and $\sqrt{s}=13$ TeV respectively. A value of $\varphi=0$ therefore corresponds to a back-to-back configuration for the $J/\psi$ and the jet and values of $\langle \cos \varphi \rangle$ close to unity are equivalent to a strong correlation. One can see from these figures that the values of $\langle \cos \varphi \rangle$ obtained with the three production mechanisms are compatible with each other as well as with the results obtained when the $J/\psi$ vertex is replaced by the leading order jet vertex shown for comparison. 
We note that passing from $\sqrt{s}=8$ TeV to $\sqrt{s}=13$ TeV increases very slightly the decorrelation effects.

One should note that these results could be significantly altered when taking into account the NLO corrections to the $J/\psi$ production vertex, as it is the case when passing from the LO to the NLO jet vertex, see refs.~\cite{Colferai:2010wu,Ducloue:2013hia}. The derivation of the NLO $J/\psi$ production vertex goes well beyond the scope 
of this 
work and is left for further studies.


\section{Conclusions}

In the present article, we have shown that the study of 
the inclusive production of a forward $J/\psi$ and a very backward jet at the LHC leads to very promising cross sections, to be studied either at the ATLAS or CMS experiments. The possibility of tagging a high rapidity jet on one side, and a $J/\psi$ charmonium on the other side (although with a smaller absolute rapidity), can give access to BFKL resummation effects, since the relative rapidity up to roughly 7 (and even 9 for CASTOR) is theoretically just in the appropriate kinematical range. We have computed the required matrix elements, in the NRQCD color singlet and color octet approaches, as well as in the color evaporation model. Our numerical results show that in the NRQCD approach, the color octet contribution dominates over the color singlet one, and the color evaporation model gives a prediction similar to the color octet NRQCD contribution. The study of the azimuthal correlations gives results which are very similar to the ones obtained in the Mueller-Navelet 
case (using for consistency one of the two jet vertices at LO, since the $J/\psi$ vertex is itself treated at LO).

The next stage, in order to get full NLL BFKL predictions for this process, would require to use the NLO expression for the charmonium production vertex, which has not yet been computed. This is left for future studies.

Finally, we did not include any double parton scattering contribution, which through two decorrelated BFKL ladders could lead to the same final state. In the case of Mueller-Navelet jets, some of us have shown that this contribution is rather small with respect to the single BFKL ladder contribution~\cite{Ducloue:2015jba}, except potentially for large $s$ and small jet transverse momenta. For the present process, it would thus be interesting to study this contribution in the CASTOR kinematics. This is left for future studies.

\acknowledgments

We thank Evgenii Baldin, Andrey Grabovsky, Jean-Philippe Lansberg and Hua-Sheng Shao for discussions.
The research by R.B. and L.Sz. was supported by the National Science Center, Poland, grant No. 2015/17/B/ST2/01838.
The research by B.D. was supported by the Academy of Finland, project 273464, and by the European Research Council, grant ERC-2015-CoG-681707.
This work is partially supported by
the French grant ANR PARTONS (Grant No. ANR-12-MONU-0008-01), by the COPIN-IN2P3 agreement,
by the Labex P2IO and by the Polish-French collaboration agreement
Polonium.
This work used computing resources from CSC -- IT Center for Science in Espoo, Finland.


\providecommand{\href}[2]{#2}\begingroup\raggedright\endgroup

\end{document}